\documentclass[11pt]{article}

\newlength{\vshift}
\newlength{\hshift}
\usepackage{amssymb,amsopn}

\renewcommand{\theequation}{\thesection.\arabic{equation}}
\newcommand{\initiate}{\setcounter{equation}{0}}

\def\nn{\nonumber }

\def\be{\beta}
\def\a{\alpha}

\def\g{\gamma}

\def\ds{\stackrel{\star}{,}}

\def\tr{{\rm Tr}}

\def\kbar{{\mathchar'26\mkern-9muk}}

\def\nn{\nonumber}
\def\be{\begin{equation}}             \def\ee{\end{equation}}
\def\ba#1{\begin{array}{#1}}          \def\ea{\end{array}}
\def\bea{\begin{eqnarray} }           \def\eea{\end{eqnarray} }
\def\beann{\begin{eqnarray*} }        \def\eeann{\end{eqnarray*} }
\def\beal{\begin{eqalign}}            \def\eeal{\end{eqalign}}
             
\def\bsubeq{\begin{subequations}}     \def\esubeq{\end{subequations}}
\def\bitem{\begin{itemize}}           \def\eitem{\end{itemize}}
                    
\def\pa{\partial}
\def\a{\alpha}
\def\b{\beta}

\def\d{\delta}
\def\g{\gamma}

\def\k{\kappa}
\def\l{\lambda}
\def\m{\mu}
\def\n{\nu}
\def\o{\omega}
\def\r{\rho}                    
\def\s{\sigma}                  

\def\G{\Gamma}

\def\L{\Lambda}

\begin{document}
\vspace{1.5cm}
\begin{titlepage}

\begin{center}

{\LARGE{\bf NC $SO(2,3)_\star$ gravity: noncommutativity as a source of curvature and torsion}}

\vspace*{1.3cm}

{{\bf Marija Dimitrijevi\' c \' Ciri\'c, Biljana Nikoli\'c  and
Voja Radovanovi\' c}}

\vspace*{1cm}

University of Belgrade, Faculty of Physics\\
Studentski trg 12, 11000 Beograd, Serbia \\[1em]

\end{center}

\vspace*{2cm}

\begin{abstract}
Noncommutative (NC) gravity is constructed on the canonical noncommutative 
(Moyal-Weyl) 
space-time as 
a noncommutative $SO(2,3)_\star$ gauge theory. The NC
gravity action
consists of three different terms: the first term is of Mac-Dowell 
Mansouri type, while the
other two are generalizations of the Einstein-Hilbert action and the 
cosmological constant
term. The expanded NC gravity action is then calculated using the Seiberg-Witten 
(SW) map and the
expansion is done up second order in the deformation parameter. We analyze in 
details the 
low energy sector of the full model. We calculate the equations of motion, 
discuss their general properties and present one solution: the NC correction to
Minkowski space-time. Using this solution, we explain breaking of the 
diffeomorphism symmetry as a consequence of working in a particular coordinate 
system given by the Fermi normal coordinates.

\end{abstract}
\vspace*{1cm}


\vspace*{1cm}
\quad\scriptsize{eMail:
dmarija@ipb.ac.rs, biljana@ipb.ac.rs and rvoja@ipb.ac.rs}
\vfill

\end{titlepage}

\setcounter{page}{1}
\newcommand{\Section}[1]{\setcounter{equation}{0}\section{#1}}
\renewcommand{\theequation}{\arabic{section}.\arabic{equation}}

\section{Introduction}

\noindent General Relativity (GR) and Quantum Field Theory (QFT) are leading 
theories in modern physics. General Relativity successfully describes gravity 
phenomena from millimeter scale to cosmic scale \cite{GRWaves}. On the other 
hand, 
Quantum Field Theory remarkably well describes physics at scales from atomic to 
elementary particle scale \cite{CERN}.
However, a theory that unites 
these two theories and provides
a description of gravity at quantum scales is still missing. One attempt to 
construct such a theory is the approach of noncommutative (NC) geometry and 
noncommutative space-time.

\noindent During the last twenty years there has been an ongoing effort to try 
to construct consistent NC gravity models. These models rely on 
the notion of NC space-time and/or noncommutative geometry and in a certain 
limit they reduce to General Relativity. One of the main problems in this 
approach is breaking of diffeomorphism symmetry of General Relativity. 
Namely, in most of NC gravity models the diffeomorphism symmetry, or at 
least a part of it, is broken and one 
needs to understand this breaking and the remaining symmetries (if any). In the 
following we mention some models of NC gravity. NC gravity via the twist 
approach 
\cite{TwistApproach} is based on the twisted diffeomorphism symmetry. One can 
write NC Einstein-Hilbert action, derive equations of motion and analyze some 
particular solutions based on the Killing or semi-Killing 
twist \cite{TwistSolutions}. However, the full meaning of the twisted symmetry 
remains to be understood better \cite{Chaichian}.
In emergent NC gravity models dynamical quantum geometry arises from NC gauge 
theory given by Yang-Mills
matrix models \cite{EmGravityApproach}. There are also fuzzy space 
gravity models and
DFR models \cite{OtherApproaches}. Finally, NC gravity can be formulated as a 
NC 
gauge theory of Lorentz or (A)dS group using the enveloping algebra approach 
and the Seiberg-Witten
(SW) map \cite{SWmapApproach, PL09}. In this approach fermions are easily 
coupled to gravity and it is straightforward to formulate NC supergravity 
models \cite{PLSUGRA}. Recently, the SW map approach was related to NC 
gravity 
models via the Fedosov deformation quantization of endomorphism bundles 
\cite{Dobrski}. There are also attempts to relate NC gravity models with some 
testable GR results like gravitational waves, cosmological solutions, 
Newtonian 
potential \cite{Ostali}.

\noindent In this article we construct a NC gravity model following the NC 
gauge theory approach. We work with
the canonical (Moyal-Weyl, $\theta$-constant) noncommutative space-time. 
However, the 
model can be straightforwardly generalized to an arbitrary NC space-time 
coming from an Abelian twist. The main disadvantage 
of the canonical NC space-time
is that, by introducing a constant NC parameter we explicitly break the
diffeomorphism symmetry.
Therefore, it is natural to ask if this symmetry breaking has some physical 
explanation. In Section 5 we will provide an explanation of this 
diffeomorphism breaking. The gauge group of our model is chosen to be the 
NC $SO(2,3)_\star$ group. Motivated by different $f(R)$, $f(T)$ and other 
modified gravity models we study
the SW map expansion of our model and obtain correction terms that are of 
the first, second, third and fourth order in powers of curvature and 
torsion. 
Those terms can be compared with the existing
terms in modified gravity models. An advantage of our model is that the 
relations between different correction terms
are not arbitrary but are fixed by the SW map expansion. Calculating NC gravity equations of
motion, we show that noncommutativity is a source of the curvature 
and torsion. That
is, given a flat/torsion-free space-time, noncommutativity induces nonzero curvature/torsion on
this space-time. This result is not completely new, it was also discussed in \cite{MajaJohn} in a
different approach to NC gravity. Especially, starting from Minkowski 
space-time as a solution of
commutative
vacuum Einstein equations, the corrections induced by our NC gravity model lead
to space-time with
a constant scalar curvature. Note that this article 
is a longer and detailed version of \cite{UsLetter}.

\noindent The structure of the article is as follows: In the following section 
we introduce the full commutative action. For completeness,
we repeat the basic notations from our previous papers \cite{MiAdSGrav}, 
\cite{MDVR-14}. After that, the full model 
consisting of a
sum of three different actions is presented. The actions are a 
MacDovell-Mansouri type of action, a generalization of the Einstein-Hilbert 
action and the cosmological constant action \cite{Wilczek}. The NC 
generalization 
of this model 
is done in
Section 3. Using the SW map the second order expansion (in the deformation 
parameter) of the
NC gravity action is calculated. The calculations are long and tedious, so we do not go into
details. Instead, we
give some of the details in Appendix B. In the zeroth order the NC action 
reduces to the
commutative action containing the Gauss-Bonnet term, Einstein-Hilbert term and the cosmological
constant term. The first order correction vanishes, as expected. The 
first non-vanishing correction is the second order 
correction. It is given by the terms that are higher order in the
curvature and torsion. Since the full second order correction is very 
complicated, in this paper we only discuss the low energy 
limit. Therefore, in Section 4 we write the expanded action keeping 
terms 
that are of zeroth, first and second order in the derivatives of vierbeins. 
The equations of motion are then obtained by varying the action
with respect to the vierbeins and the spin-connection. NC corrections
($\theta$-dependent terms) appear on the right-hand side of these 
equations and can be interpreted as sources of curvature and/or torsion. Using 
these equations of motion, in Section 5 we calculate the NC correction
to Minkowski space-time. We see that due to the 
noncommutativity, Minkowski
space-time becomes curved with a constant scalar curvature and the full metric 
is very close in form to the
metric of the AdS space-time. The coordinates in which the solution is given turn out to be
Fermi-normal coordinates. This result, its relation with the diffeomorphism 
symmetry breaking and the work in perspective we discuss in the Conclusions.

\initiate
\section{Commutative model}

\noindent In this section we review the commutative model. We first repeat the 
basic notation
and then define and discuss the commutative action.

\noindent Let us consider a gauge theory on four dimensional Minkowski 
space-time with 
the $SO(2,3)$ group as the gauge group. Note that through the paper we use the 
"mostly minus"
convention for the metric, $\eta_{\mu\nu}={\rm
diag}(+,-,-,-,)$. See Appendix A for more details on the conventions we use. The
gauge field is valued in the $SO(2,3)$ algebra
\be
\omega_\m = \frac{1}{2}\omega_\m^{AB}M_{AB} ,\label{GaugePotAds}
\ee
where $M_{AB}$ are the generators of the $SO(2,3)$ group. The generators 
satisfy 
\be
[M_{AB},M_{CD}]=i(\eta_{AD}M_{BC}+\eta_{BC}M_{AD}-\eta_{AC}M_{BD}-\eta_{BD}M_{AC
}) .
\label{AdSalgebra}
\ee
The $5 D$ metric is $\eta_{AB}={\rm diag}(+,-,-,-,+)$. The gauge group indices 
$A,B,\dots$ take
values $0,1,2,3,5$, while indices $a,b,\dots$ take values $0,1,2,3$. The 
space-time indices
we label with Greek letters. The generators $M_{AB} $ can be defined using the 
Clifford algebra in 5D.
A representation of 5D gamma matrices is obtained from 4D gamma matrices, i. e. 
$\Gamma_A =(i\gamma_a\gamma_5, \gamma_5)$, where $\g_a$ are 4D gamma matrices. 
The generators $M_{AB}$ are
\be 
M_{AB} =\frac{i}{4}[\Gamma_A,\Gamma_B]\ .
\ee 
In the representation given above we obtain
\bea
M_{ab} &=&\frac{i}{4}[\gamma_a,\gamma_b]=\frac12\sigma_{ab}\ ,\nonumber\\
M_{5a} &=&\frac{1}{2}\gamma_a\ . \label{Maba5}
\eea
Using this representation, the
gauge filed $\omega_\m^{AB}$ can be decomposed as:
\be
\omega_\m =
\frac{1}{2}\omega_\m^{AB}M_{AB}=\frac{1}{4}\omega_\m^{ab}\sigma_{ab}-
\frac{1}{2l} e_\m^{a}\gamma_a .\label{GaugePotAdsDecomp}
\ee
The parameter $l$ has dimension of length, while fields $e_\mu^a$ are 
dimensionless and
$\omega_\m^{ab}$ has dimension $1/l$. Under the $SO(2,3)$ gauge
transformations the gauge field transforms as
\be
\delta_\epsilon\omega_\mu=\pa_\mu\epsilon + i[\epsilon, \omega_\m],
\label{TrLawOmegaAB}
\ee
with the gauge parameter denoted by $\epsilon=\frac{1}{2}\epsilon^{AB}M_{AB}$.

\noindent The field strength tensor is defined in the standard way as
\be
F_{\m\n}=\pa_\m\omega_\n-\pa_\n\omega_\m-i[\omega_\m,\omega_\n]
=\frac{1}{2}F^{AB}_{\mu\nu}M_{AB} . \label{FAB}
\ee
Its transformation law under the infinitesimal gauge transformations is given 
by 
\be
\delta_\epsilon F_{\mu\nu}= i[\epsilon, F_{\m\nu}]. \label{TrLawFmini}
\ee
Just like the gauge potential, the components of the field strength tensor
$F_{\m\n}^{\ AB}$ decompose into $F_{\m\n}^{\ ab}$ and $F_{\m\n}^{\ a5}$:
\be
F_{\m\n}=\frac12\Big( R_{\m\n}^{\ 
ab}-\frac{1}{l^2}(e_\m^ae_\n^b-e_\m^be_\n^a)\Big)
M_{ab} + F_{\m\n}^{\ a5}M
_{a5} , \label{FabFa5}
\ee
where
\bea
R_{\m\n}^{\ ab} &=&
\pa_\m\omega_\n^{ab}-\pa_\n\omega_\m^{ab}+\omega_\m^{ac}\omega_\n^{cb}
-\omega_\m^{bc}\omega_\n^{ca} , \label{Rab}\\
lF_{\m\n}^{\ a5} &=& \nabla_\m e^a_\n-\nabla_\n e^a_\m = T_{\m\n}^a .\label{Ta}
\eea 
The $SO(2,3)$ gauge theory was used in \cite{stelle-west} to formulate a gravity 
theory using
the symmetry breaking from $SO(2,3)$ to $SO(1,3)$. Then, using the equations of 
motion of the model one can identify the fields
$\omega_\mu^{ab}$ with the spin connection and the fields $e^a_\m$ 
with vierbeins. The fields
strengths $R_{\m\n}^{\ ab}$ and $F_{\m\n}^{\ a5}=T_{\m\n}^a$ are the curvature 
tensor and the
torsion.

\noindent The symmetry breaking was introduced via the scalar
field $\phi=\phi^A\Gamma_A$ which 
transforms in the
adjoint representation of the $SO(2,3)$ group,
\be
\delta \phi = i[\epsilon,\phi] . \label{DeltaPhi}
\ee
Using the scalar field $\phi$ one can write the following gauge invariant 
actions \cite{Wilczek}:
\begin{eqnarray}
S_1 &=& \frac{il}{64\pi G_N}\tr \int{\rm d}^4x \epsilon^{\mu\nu\rho\sigma}
F_{\mu\nu} F_{\rho\sigma}\phi  ,\label{KomDejstvo_S_1}\\
S_2 &=& \frac{1}{128 \pi G_{N}l}\tr\int d^{4}x\epsilon^{\mu \nu \rho 
\sigma}F_{\mu 
\nu}D_{\rho}\phi D_{\sigma}\phi\phi+c.c. , \label{KomDejstvo_S_2}\\
S_3 &=& -\frac{i}{128 \pi G_{N}l}\tr\int d^{4}x\epsilon^{\mu \nu \rho 
\sigma}D_{\m}\phi D_{\n}\phi D_{\rho}\phi D_{\sigma}\phi\phi 
,\label{KomDejstvo_S_3} 
\end{eqnarray}
with $D_\mu\phi = \partial_\mu\phi -i[\omega_\mu, \phi]$. 

\noindent We define our commutative model to be the sum of these three actions
\begin{equation}
S=c_1S_1+c_2S_2+c_3S_3 , \label{FullCommAction}
\end{equation}
where $c_1,c_2$ and $c_3$ are arbitrary constants that will be determined 
from
some additional constraints. The action (\ref{FullCommAction}) is invariant 
under the $SO(2,3)$
gauge symmetry. This symmetry is broken to the $SO(1,3)$ gauge symmetry by 
choosing
$\phi^a=0,\phi^5=l$. This choice is sometimes referred to as a physical 
gauge. After the symmetry breaking the 
action $S_1$ reduces to the sum of the Einstein-Hilbert term, the cosmological 
constant term
and the Gauss-Bonnet term:
\begin{equation}
S_1 = -\frac{1}{16\pi
G_N}\int {\rm d}^4 x\Big( \frac{l^2}{16}\epsilon^{\m\n\r\s}
\epsilon_{abcd}R_{\m\n}^{\ ab}R_{\r\s}^{\ cd} + eR -\frac{6}{l^2} e \Big) .
\nn
\end{equation}
The action $S_2$ reduces to the sum of the Einstein-Hilbert term and the 
cosmological constant term
\begin{equation}
S_2=-\frac{1}{16\pi G_{N}}\int 
d^{4}x\sqrt{-g}\Big(R-\frac{12}{l^2}\Big) .\nonumber
\end{equation}
Finally, the action $S_3$ reduces to the cosmological constant term only
\begin{equation}
S_3=-\frac{1}{16\pi G_{N}}\int 
d^{4}x\sqrt{-g}\Big(-\frac{12}{l^2}\Big) .
\end{equation}
Therefore, after the symmetry breaking our classical action is a sum of these 
three terms
\bea 
S&=&c_1S_1+c_2S_2+c_3S_3\nn\\
&=&-\frac{1}{16\pi G_{N}}\int 
d^{4}x\Big(c_1\frac{l^2}{16}\epsilon^{\m\n\r\s}
\epsilon_{abcd}R_{\m\n}^{\ ab}R_{\r\s}^{\ cd} \nn\\
&& +\sqrt{-g}\big( (c_1 + c_2)R -\frac{6}{l^2}(c_1+ 2c_2 + 2c_3)\big)
\Big). \label{KomDejstvo} 
\eea
Now we can partially fix the constants $c_1$, $c_2$ and $c_3$
by the requirement that the full action after the symmetry breaking reduces to 
the Einstein Hilbert action with the cosmological constant. The Gauss-Bonet term 
is topological, it
does not influence the equations of motion and we will not write it further. 
We choose $c_1+c_2=1$, and the cosmological constant is given by 
$$\L=-3\frac{1+c_2+2c_3}{l^2}\ . $$ 
Note that the cosmological constant $\L$ can be positive, negative or 
zero, regardless of the symmetry of our model. 

\initiate
\section{NC $SO(2,3)_\star$ gravity action}

\noindent As we have mentioned in Introduction, the NC generalization of 
General 
Relativity cannot be formulated in a straightforward way. One 
of possible ways to 
achieve this is to
use knowledge of the NC gauge theories and treat gravity as a gauge theory of 
the Poincar\'e (or
AdS or dS) group. In the previous section we defined a rather general model 
of commutative
gravity as a theory with broken $SO(2,3)$ symmetry (\ref{KomDejstvo}). This 
model we now
generalize to the noncommutative setting.

\noindent As in the previous papers \cite{MiAdSGrav, MDVR-14}, we work in the 
canonical (Moyal-Weyl, $\theta$-constant) NC
space-time. Following the approach
of deformation quantization we represent noncommutative functions as functions
of commuting coordinates and algebra multiplication with the Moyal-Weyl
$\star$-product: 
\begin{equation}
\label{moyal}  f (x)\star  g (x) =
      e^{\frac{i}{2}\,\theta^{\a\b}\frac{\pa}{\pa x^\a}\frac{\pa}{ \pa
      y^\b}} f (x) g (y)|_{y\to x}\ .
\end{equation}
Here $\theta^{\a\b}$ is a constant antisymmetric matrix and its entries are 
considered to
be small deformation parameters\footnote{To be more precise, the Moyal-Weyl 
$\star$-product should be written as
$$
f (x)\star  g (x) =
      e^{\frac{i}{2}\,\kbar\theta^{\a\b}\frac{\pa}{\pa x^\a}\frac{\pa}{ \pa
      y^\b}} f (x) g (y)|_{y\to x}\ ,
$$      
with the small deformation parameter $\kbar$ and arbitrary constant 
antisymmetric matrix elements $\theta^{\a\b}$. In the usual notation $\kbar$ is 
absorbed in the matrix elements $\theta^{\a\b}$ and these are called 
deformation parameters.
}. The noncommutativity (deformation) is 
then encoded in
the $\star$-product, while all
variables (fields) are functions of commuting coordinates. Integration is well
defined since the usual integral is cyclic:
\be
\int {\rm d}^4 x (f\star g\star h ) = \int {\rm d}^4 x ( h\star f\star g )\
 + {\mbox{boundary terms}}.\label{cyclicity}
\ee
Assuming that all fields are well behaved at the boundary, these terms 
vanish and since
we are interested in the equations of motion, we will simply ignore the boundary 
terms
throughout this paper. They become important when one discuss conserved 
quantities or
thermodynamics of black holes. The question of boundary terms in the NC gravity 
action we
discussed in details in \cite{MDVR-14}. In particular, from (\ref{cyclicity}) we 
have $\int {\rm
d}^4 x (f\star g) = \int {\rm d}^4 x (g\star f) =\int
{\rm d}^4 x fg$. Note that the volume element ${\rm d}^4 x$ is not 
$\star$-multiplied
with functions under the integral.

\noindent In order to construct the NC $SO(2,3)_\star$ gauge theory we use the
enveloping algebra approach
and the Seiberg-Witten map developed in \cite{SWMapEnvAlgebra}. We will not 
go into
details of this construction, they can be found in \cite{MDVR-14}. Here we 
just write the SW
map solutions for the NC gauge field,  NC field strength tensor and the NC 
scalar field in
the adjoint representation since we will use them through the paper.

\noindent The noncommutative gauge
field $\hat{\omega}_\m$ is defined by the following recursive 
relation:
\be
{\hat\omega}_\m^{(n+1)}= -\frac{1}{4(n+1)}\theta^{\kappa\lambda}
\Big( \{{\hat \omega}_\kappa \ds \pa_\lambda{\hat \omega}_\m + {\hat F}_{\l\m}\}
\Big)^{(n)} ,\label{RecRelOmega}
\ee
where $\hat{\omega}^{(0)}_\m = \omega_\mu$ is the commutative gauge field 
and an expression of the type $(A\star B)^{(n)} = 
A^{(n)}B^{(0)} + A^{(n-1)}B^{(1)} + 
\dots
+ A^{(0)}\star ^{(1)} B^{(n-1)} + A^{(1)}\star ^{(1)} B^{(n-2)} +\dots$ includes
all possible terms of order $n$. Expanding this relation up to first order 
in the deformation parameter, we find that the
NC gauge field ${\hat\omega}_\m$ is of the form
\bea
{\hat\omega}_\m &=& \omega_\mu
-\frac{1}{4}\theta^{\kappa\lambda}\{\omega_\kappa, (\partial_\lambda\omega_\mu +
F_{\lambda\mu}\}
+ {\cal O}(\theta^2) \label{SO23Omega1}\\
&=& \frac{1}{4}\omega^{ab}_\m\sigma_{ab}  + \omega_\mu^a \gamma_a
+\tilde{\omega}_\mu^a\gamma_a\gamma_5 
+ {\tilde\omega}^5_\m\gamma_5 + \omega_\m I \ . \label{UEAOmega}
\eea
It is obvious from (\ref{UEAOmega}) that the NC gauge field is valued in the 
enveloping algebra of
the $SO(2,3)$ algebra. However, note that the enveloping algebra in this 
particular case is finite
dimensional. This is one of the advantages of choosing the NC gauge group to be 
$SO(2,3)_\star$.

\noindent The NC field strength tensor is defined as
\be
{\hat F}_{\m\n}=\pa_\m{\hat \omega}_\n-\pa_\n{\hat \omega}_\m
-i[{\hat\omega}_\m\ds {\hat\omega}_\n] \label{nckrivina}
\ee
and its transformation law under the infinitesimal NC gauge transformations is 
given by:
\begin{equation}
\delta^\star_\epsilon {\hat F}_{\m\n}= i[\hat{\Lambda}_\epsilon \ds {\hat
F}_{\m\n}]\ .\label{DeltaFStar}
\end{equation}
Here the NC gauge parameter $\hat{\Lambda}_\epsilon$ is introduced. It is also 
valued in the
enveloping algebra, in zeroth order in the deformation parameter it reduces 
to the commutative
$SO(2,3)$ gauge parameter $\epsilon$ and its higher orders can be calculated 
using the SW map. The
SW map solution for ${\hat F}_{\mu\nu}$ follows from
the definition (\ref{nckrivina}), using the result (\ref{RecRelOmega}). The
recursive formula is
\bea
{\hat F}_{\m\n}^{(n+1)} &=& -\frac{1}{4(n+1)}\theta^{\kappa\lambda}\Big( \{
{\hat \omega}_\kappa \ds
\partial_\lambda {\hat F}_{\mu\nu} + D_\lambda {\hat F}_{\mu\nu} \} \Big)^{(n)}
\nn\\
&& +\frac{1}{2(n+1)}\theta^{\kappa\lambda}\Big( \{ {\hat F}_{\mu\kappa}, \ds
{\hat F}_{\nu\lambda} \}
\Big)^{(n)} \ .\label{RecRelR}
\eea
Note that we do not put a "hat" on the covariant derivative $D_\mu$, the meaning
of $D_\mu$ is defined by the expression it acts on: $D_\lambda \hat{F}_{\m\n} =
\partial_\lambda \hat{F}_{\mu\nu}
-i[\hat{\omega}_\lambda \ds \hat{F}_{\mu\nu}]$ and $D_\lambda F_{\m\n} =
\partial_\lambda F_{\mu\nu}
-i[\omega_\lambda, F_{\mu\nu}]$. One can check that
\bea
{\hat F}_{\mu\nu} &=& F_{\mu\nu} -\frac{1}{4}\theta^{\kappa\lambda}
\{\omega_\kappa,\pa_\lambda F_{\m\n} + D_\lambda F_{\m\n}\} +
\frac{1}{2}\theta^{\kappa\lambda} \{F_{\mu\kappa}, F_{\nu\lambda} \}
+ {\cal O}(\theta^2) \label{SO23F1}\\
&=& \frac{1}{4} F^{\ ab}_{\m\n}\sigma_{ab} + F^a\gamma_a +
\tilde{F}^a\gamma_a\gamma_5
+ {\tilde  F}^5_{\m\n}\gamma_5  + F_{\m\n} I . \label{UEAF}
\eea

Finally, the field $\hat\phi$ transforms in the adjoint representation 
\be
\delta_\epsilon^\star{\hat \phi} = i[{\hat\Lambda}_\epsilon\ds{\hat \phi}]\  .
\label{DeltaPhiStar}
\ee
Using the previous results we find the recursive relation
\be
\hat{\phi}^{(n+1)} = -\frac{1}{4(n+1)}\theta^{\kappa\lambda} \Big( \{{\hat
\omega}_\kappa \ds \pa_\l {\hat{\phi}} + D_\l {\hat{\phi}} \} \Big)^{(n)} \ ,
\label{RecRelPhi}
\ee
with $D_\l {\hat \phi} = \partial_\l {\hat \phi} -i [{\hat \omega}_\l \ds {\hat 
\phi} ]$ and $D_\l \phi = \partial_\l \phi -i [\omega_\l ,\phi ]$.
The solution for $\hat{\phi}$ has the following structure
\bea
{\hat \phi} &=& \phi -\frac{1}{4}\theta^{\kappa\lambda}
\{\omega_\kappa,\pa_\lambda\phi + D_\lambda \phi\} + {\cal O}(\theta^2)
\label{SO23Phi1}\\
&=& \phi^a\gamma_a\gamma_5 + \phi\gamma_5 + \frac{1}{4}\phi^{ab}\sigma_{ab} +
\tilde{\phi}^a\gamma_a\ . \label{UEAPhi}
\eea

\noindent Having these results at hand, we are now ready to define a NC 
generalization of
the action (\ref{KomDejstvo}). We do it term by term.

\subsection{NC generalization of $S_1$}

\noindent The NC generalization of the action $S_1$ (\ref{KomDejstvo_S_1}) is 
given by
\be
S_{1NC} = \frac{il}{64\pi G_N}\tr \int{\rm d}^4x \epsilon^{\mu\nu\rho\sigma}
\hat{F}_{\mu\nu}\star \hat{F}_{\rho\sigma}\star \hat{\phi}\, 
.\label{NCdejstvo_S_1}
\ee
The $\star$-product is the Moyal-Weyl $\star$-product (\ref{moyal}),
fields with a "hat" are NC fields and we will use the SW map solutions
(\ref{SO23F1}), (\ref{SO23Phi1}). Using the transformation laws
(\ref{DeltaFStar}), (\ref{DeltaPhiStar}) and the cyclicity of the integral
(\ref{cyclicity}) one can show that this action is invariant under the NC
$SO(2,3)_\star$ gauge transformations. In the limit $\theta^{\alpha\beta}\to 0$
the
action (\ref{NCdejstvo_S_1}) reduces to the commutative action 
(\ref{KomDejstvo_S_1}).

\noindent The expansion of this action up the the second order in the 
deformation parameter
is done in \cite{MDVR-14}. The first order correction vanishes. This is an 
expected result: it was shown in \cite{SWmapApproach} that, if the NC gravity 
action is
real, then the first order (in the deformation parameter) correction has to 
vanish. This result
holds for a wide class of NC deformations, namely the deformations obtained by 
an Abelian twist, see \cite{PL09}. The second 
order correction is given by:
\bea 
S_{1NC}^{(2)} &=& \frac{il}{64\pi
G_N}\frac18\theta^{\a\b}\theta^{\kappa\l}\tr \int{\rm d}^4x
\epsilon^{\mu\nu\rho\sigma}\Big( \frac18\{
F_{\a\b},\{F_{\m\n},F_{\r\s}\}\}\{\phi,F_{\kappa\l}\} \nn\\
&& -\frac12\{F_{\a\b},\{F_{\r\s},\{F_{\kappa\m},F_{\l\n}\}\}\}\phi
-\frac14\{\{F_{\m\n},F_{\r\s}\},\{F_{\kappa\a},F_{\l\b}\}\}\phi \nn\\
&& -\frac{i}{4}\{F_{\a\b},[D_\kappa
F_{\m\n},D_{\l}F_{\r\s}]\}\phi - \frac{i}{2}[\{D_\kappa
F_{\m\n},F_{\r\s}\},D_{\l}F_{\a\b}]\phi \nn\\
&& -\frac12\{
F_{\r\s},\{F_{\a\m},F_{\b\n}\}\}\{\phi,F_{\kappa\l}\}
+\{ \{F_{\a\m},F_{\b\n} \},\{F_{\kappa\r},F_{\l\s}\}\}\phi\nn\\
&& +2\{F_{\r\s},\{F_{\b\n},\{F_{\kappa\a},F_{\l\m}\}\}\}\phi
+ i\{F_{\r\s},[D_\kappa F_{\a\m},D_\l F_{\b\n}]\}\phi \nn\\
&& +2i[\{F_{\b\n},D_{\kappa}F_{\a\m}\},D_\l F_{\r\s}]\phi \nn\\
&&-\frac{i}{4}\{\phi,F_{\kappa\l}\}[D_\a F_{\m\n},D_\b F_{\r\s}]
-\frac{1}{2}\{D_\kappa D_\a F_{\m\n},D_\l D_\b F_{\r\s}\}\phi\nn\\
&& + i[\{ F_{\k\a},D_\l F_{\m\n}\},D_\b F_{\r\s}]\phi
+ i[\{ F_{\l\n},D_\a F_{\kappa\m}\},D_\b F_{\r\s}]\phi \nn\\
&& + i[\{ F_{\k\m},D_\a F_{\l\n}\},D_\b F_{\r\s}]\phi\Big) \ .
\label{NCDejstvo1Exp2}
\eea

\subsection{NC generalization of $S_2$}

\noindent The NC generalization of the action $S_2$ (\ref{KomDejstvo_S_2}) is 
given by
\begin{equation} 
S_{2NC}=\frac{1}{128 \pi G_{N}l}\tr \int d^{4}x \epsilon^{\mu \nu 
\rho \sigma}\hat\phi\star\hat F_{\mu \nu}\star\hat D_{\rho}\hat\phi\star\hat 
D_{\sigma}\hat\phi + c.c. \label{NCDejstvo_S_2} 
\end{equation}
This action is not real so we have to add its complex 
conjugate by hand. Following
the usual steps, we expand (\ref{NCDejstvo_S_2}) up to second order in the 
deformation
parameter. The details of the calculation are presented in Appendix B, here 
we just write
the main steps.

\noindent Using the formulae (\ref{B1}), (\ref{B2}) and (\ref{B3}) from 
Appendix B the first
order correction follows. It is given by
\begin{eqnarray}
S_{2NC}^{(1)}&=& \frac{1}{128 \pi G_{N}l}\tr\int d^{4}x \theta^{\alpha 
\beta}\epsilon^{\mu \nu \rho 
\sigma}\Big( -\frac{1}{4}\phi\{F_{\alpha \beta},F_{\mu \nu}\}D_{\rho} \phi 
D_{\sigma}\phi\nn\\
&& -\frac{i}{2}D_{\alpha}\phi F_{\mu 
\nu}(D_{\beta}D_{\rho}\phi)D_{\sigma}\phi -\frac{i}{2}D_{\alpha}\phi 
F_{\mu \nu}D_{\rho}\phi (D_{\beta}D_{\sigma}\phi)+\nonumber\\
&&+\frac{1}{2}\phi\{F_{\mu \alpha},F_{\nu \beta}\}D_{\rho}\phi 
D_{\sigma}\phi+\frac{i}{2}\phi F_{\mu 
\nu}(D_{\alpha}D_{\rho}\phi)(D_{\beta}D_{\sigma}\phi)\nn\\&+&\frac{1}{2}\phi 
F_{\mu \nu}\{F_{\alpha \rho},D_{\beta}\phi\}D_{\sigma}\phi+\frac{1}{2}\phi 
F_{\mu \nu}D_{\rho}\phi\{F_{\alpha \sigma},D_{\beta}\phi\} \Big) + c.c.
\end{eqnarray}
Explicit calculation of traces gives $S^{(1)}_2=0$, so we have to calculate 
the second order correction.  It follows from the first order action as
\begin{eqnarray}
S_{2NC}^{(2)}&=& \frac{1}{256 \pi G_{N}l}\tr\int d^{4}x \theta^{\alpha 
\beta}\epsilon^{\mu \nu \rho 
\sigma}Tr\Big(-\frac{1}{4}\phi\{\hat {F}_{\alpha \beta}\ds\hat {F}_{\mu 
\nu}\}\star \hat D_{\rho} \hat\phi 
\star \hat D_{\sigma}\phi\nn\\
&&-\frac{i}{2}\hat D_{\alpha}\phi\star  \hat 
{F}_{\mu\nu}\star (\hat D_{\beta}\hat 
D_{\rho}\hat \phi)\star \hat D_{\sigma}\hat \phi-\frac{i}{2}\hat 
D_{\alpha}\hat \phi \star \hat {F}_{\mu 
\nu}\star \hat D_{\rho}\hat \phi \star (\hat D_{\beta}\hat D_{\sigma}\hat 
\phi)\label{S_2-drugi-red-1}\\
&&+\frac{1}{2}\hat \phi\star\{\hat {F}_{\mu \alpha}\ds\hat {F}_{\nu 
\beta}\}\star\hat D_{\rho}\hat \phi 
\star\hat D_{\sigma}\hat \phi+\frac{i}{2}\hat \phi \star\hat {F}_{\mu 
\nu}\star(\hat D_{\alpha}\hat 
D_{\rho}\hat \phi)\star(\hat D_{\beta}\hat D_{\sigma}\hat \phi)\nn\\
&&+\frac{1}{2}\hat \phi\star 
\hat {F}_{\mu 
\nu}\star\{\hat {F}_{\alpha \rho}\ds\hat 
D_{\beta}\hat \phi\}\star\hat D_{\sigma}\hat \phi+\frac{1}{2}\hat \phi\star 
\hat 
{F}_{\mu\nu}\star\hat D_{\rho}\hat \phi\{\hat {F}_{\alpha \sigma}\ds\hat 
D_{\beta}\hat \phi\}\Big)^{(1)} + c.c.\nn
\end{eqnarray}
By $()^{(1)}$ it is meant the terms in the bracket are expanded up to
first order in the
deformation parameter. That includes expansion of the $\star$-products and the 
use of the SW
map solutions for the corresponding fields. 

\noindent Using the formulae (\ref{B4}-\ref{B6}) and the general method outlined 
in
Appendix B, we finally arrive at the second order correction for the NC action 
$S_2$
\begin{eqnarray}
S^{(2)}_{2NC}&=&\frac{1}{256\pi G_{N}l}\int d^{4}x \epsilon^{\mu \nu 
\rho \sigma}\theta^{\alpha \beta}\theta^{\gamma 
\delta}\tr\Bigg(\frac{1}{8}\{F_{\gamma \delta},\phi\{F_{\alpha 
\beta},F_{\mu \nu}\}\}D_{\rho}\phi D_{\sigma}\phi\nn\\
&&-\frac{i}{4}D_{\gamma}\phi 
D_{\delta}(\{F_{\alpha \beta},F_{\mu \nu}\})D_{\rho}\phi D_{\sigma}\phi 
\nonumber\\
&& -\frac{i}{4}\phi [D_{\gamma}F_{\alpha \beta}, D_{\delta}F_{\mu 
\nu}]D_{\rho}\phi D_{\sigma}\phi-\frac{1}{4}\phi\{\{F_{\alpha \gamma},F_{\beta 
\delta}\},F_{\mu \nu}\}D_{\rho}\phi D_{\sigma}\phi\nonumber\\
&& -\frac{1}{4}\phi\{\{F_{\mu \gamma},F_{\nu \delta}\},F_{\alpha 
\beta}\}D_{\rho}\phi D_{\sigma}\phi-\frac{i}{4}\phi\{F_{\alpha \beta},F_{\mu 
\nu}\}D_{\gamma}D_{\rho}\phi D_{\delta}D_{\sigma}\phi\nonumber\\
&& -\frac{1}{4}\phi\{F_{\alpha \beta},F_{\mu \nu}\}[\{F_{\gamma 
\rho},D_{\delta}\phi\},D_{\sigma}\phi]
+\frac{i}{4}\{F_{\gamma \delta},D_{\alpha}\phi F_{\mu 
\nu}\}[D_{\beta}D_{\rho}\phi,
D_{\sigma}\phi]\nn\\
&& +\frac{1}{2}(D_{\gamma}D_{\alpha}\phi) D_{\delta}F_{\mu 
\nu}[D_{\beta}D_{\rho}\phi, D_{\sigma}\phi]\nonumber\\
&& -\frac{i}{2}\{F_{\gamma \alpha},D_{\delta}\phi\}F_{\mu 
\nu}[D_{\beta}D_{\rho}\phi, D_{\sigma}\phi]-\frac{i}{2}D_{\alpha}\phi\{F_{\mu 
\gamma},F_{\nu \delta}\}[D_{\beta}D_{\rho}\phi, D_{\sigma}\phi]\nonumber\\
&& +\frac{1}{2}D_{\alpha}\phi F_{\mu 
\nu}\{D_{\gamma}D_{\beta}D_{\rho}\phi, D_{\delta}D_{\sigma}\phi\}
-\frac{i}{2}D_{\alpha}\phi F_{\mu \nu}[\{F_{\gamma 
\beta},D_{\delta}D_{\rho}\phi\},D_{\sigma}\phi]
\nonumber\\
&& -\frac{i}{2}D_{\alpha}\phi F_{\mu \nu}D_{\beta}([\{F_{\gamma 
\rho},D_{\delta}\phi\},D_{\sigma}\phi])-\frac{1}{4}\{F_{\gamma 
\delta},\phi\{F_{\mu \alpha},F_{\nu 
\beta}\}\}D_{\rho}\phi D_{\sigma}\phi \nn\\
&& +\frac{i}{2}D_{\gamma}\phi D_{\delta}(\{F_{\mu \alpha},F_{\nu 
\beta}\})D_{\rho}\phi D_{\sigma}\phi 
+\frac{i}{2}\phi [D_{\gamma}F_{\mu \alpha},D_{\delta}F_{\nu 
\beta}]D_{\rho}\phi D_{\sigma}\phi\nn\\
&& +\phi \{\{F_{\mu \gamma},F_{\alpha 
\delta}\},F_{\nu \beta}\}D_{\rho}\phi D_{\sigma}\phi\nonumber\\
&& +i\phi \{F_{\mu \gamma},F_{\nu 
\delta}\}D_{\alpha}D_{\rho}\phi D_{\beta}D_{\sigma}\phi\nonumber\\
&& +\phi\{F_{\mu 
\alpha},F_{\nu \beta}\}[\{F_{\gamma \rho},D_{\delta}\phi\},D_{\sigma}\phi]
-\frac{i}{4}\{F_{\gamma \delta},\phi F_{\mu 
\nu}\}D_{\alpha}D_{\rho}\phi D_{\beta}D_{\sigma}\phi\nonumber\\
&& +\phi F_{\mu \nu}\{F_{\alpha \rho},D_{\beta}\phi\}\{F_{\gamma 
\sigma},D_{\delta}\phi\} -\frac{1}{2}D_{\gamma}\phi D_{\delta}F_{\mu 
\nu}D_{\alpha}D_{\rho}\phi 
D_{\beta}D_{\sigma}\phi \nn\\
&& +i\phi F_{\mu \nu}\{D_{\alpha}(\{F_{\gamma 
\rho},D_{\delta}\phi\}),D_{\beta}D_{\sigma}\phi\}\nonumber\\
&& +\frac{i}{2}\phi F_{\mu \nu}\{\{F_{\gamma 
\alpha},D_{\delta}D_{\rho}\phi\},D_{\beta}D_{\sigma}\phi\}\nonumber\\
&& -\frac{1}{4}\{F_{\gamma \delta},\phi\ 
F_{\mu \nu}\}[\{F_{\alpha \rho},D_{\beta}\phi\},D_{\sigma}\phi]\nonumber\\
&& +\frac{i}{2}D_{\gamma}\phi D_{\delta}F_{\mu \nu}[\{F_{\alpha 
\rho},D_{\beta}\phi\},D_{\sigma}\phi]
+\frac{i}{2}\phi F_{\mu \nu}[[D_{\gamma}F_{\alpha 
\rho},D_{\delta}D_{\beta}\phi],D_{\sigma}\phi]\nn\\
&& -\frac{1}{2}\phi F_{\mu 
\nu}D_{\gamma}(D_{\alpha}D_{\rho}\phi)D_{\delta}(D_{\beta}D_{\sigma}
\phi)+\frac{1}{2}\phi F_{\mu 
\nu}[\{\{F_{\alpha \gamma},F_{\rho 
\delta}\},D_{\beta}\phi\},D_{\sigma}\phi]\nonumber\\
&& +\frac{1}{2}\phi F_{\mu \nu}[\{\{F_{\gamma \beta},D_{\delta}\phi\},F_{\alpha 
\rho}\},D_{\sigma}\phi] \Bigg) .\label{NCDejstvo_S2-drugi-red-2}
\end{eqnarray}

\subsection{NC generalization of $S_3$}

\noindent Finally, we consider the NC generalization of the action $S_3$ 
(\ref{KomDejstvo_S_3}).
Inserting $\star$-products and promoting the commutative fields to the 
corresponding NC
fields we arrive at: 
\begin{equation}
S_{3NC}= -\frac{i}{128 \pi G_{N}l}\tr\int \mathrm{d}^{4} x\, 
\varepsilon^{\mu\nu\rho\sigma} D_\m 
\hat\phi \star D_\n \hat\phi\star
\hat{D}_{\rho}\hat{\phi} \star \hat{D}_{\sigma}\hat{\phi}\star \hat{\phi}
.\label{NCDejstvo_S_3}
\end{equation}
The zeroth order of the action (\ref{NCDejstvo_S_3}) is the commutative action 
given by 
(\ref{KomDejstvo_S_3}).

\noindent Following the same steps as in previous subsections and using the 
formulae from
Appendix B we calculate the first order correction to this action:
\begin{eqnarray}
S_{3NC}^{(1)}&=&
-\frac{i}{128\pi G_Nl^3}\theta^{\a\b}\int \mathrm{d}^{4} 
x\epsilon^{\m\n\r\s}\tr\Big(
-\frac14\{F_{\a\b},D_\m\phi D_\n\phi\}D_\r\phi D_\s\phi \phi \nn\\
&&+\frac12\Big(\frac{i}{2}(D_\a D_\m\phi)(D_\b
D_\n\phi) + \frac12\{F_{\a\m},D_\b\phi\}D_\n\phi \nn\\
&&+\frac12D_\m\phi\{F_{\a\n}
,D_\b\phi\}\Big) D_\r\phi D_\s\phi \phi\nn\\
&& +D_\m\phi D_\n\phi\Big( \frac{i}{2}D_\a(D_\r\phi D_\s\phi)D_\b\phi
+\frac{i}{2}(D_\a D_\r\phi) (D_\b D_\s\phi)\phi\nn\\
&& +\frac12\{F_{\a\r},D_\b\phi\}
D_\s\phi \phi+\frac12D_\r \phi\{F_{\a\s},D_\b\phi\}\phi\Big)\Big) 
.\label{NCDejstvo_S_3_prvi_red}
\end{eqnarray}
Again, there is no surprise to find that the calculation of traces leads to 
$S_{3NC}^{(1)} =0$. Therefore, the first non-vanishing correction is the second 
order correction. To calculate it we start from:
\begin{eqnarray}
S_{3NC}^{(2)}&=&
-\frac{i}{256\pi G_Nl^3}\int \mathrm{d}^{4} x\epsilon^{\m\n\r\s}\tr\Big(
-\frac14\{\hat F_{\a\b}\ds \hat D_\m\hat{\phi} \hat D_\n\hat{\phi}\}\star \hat 
D_\r\hat{\phi}\star \hat D_\s\hat{\phi}\star \hat{\phi}\nn\\
&&+
\frac12\Big(\frac{i}{2}(\hat D_\a \hat D_\m\hat{\phi})\star(\hat D_\b \hat 
D_\n\hat{\phi})\nn\\
&&+\frac12\{\hat F_{\a\m}\ds \hat D_\b\hat{\phi}\}\star \hat 
D_\n\hat{\phi}+\frac12\hat D_\m\hat{\phi}\star\{\hat F_{\a\n}
\ds \hat D_\b\hat{\phi}\}\Big)\star \hat D_\r\hat{\phi}\star \hat D_\s\hat{\phi} 
\star\hat{\phi}\nn\\
&&+ \hat D_\m\hat{\phi}\star \hat D_\n\hat{\phi}\star\Big( \frac{i}{2}\hat 
D_\a(\hat D_\r\hat{\phi} \star \hat D_\s\hat{\phi})\star \hat D_\b\hat{\phi}
+\frac{i}{2}(\hat D_\a \hat D_\r\hat{\phi}) \star(\hat D_\b \hat 
D_\s\hat{\phi})\star\hat{\phi}\nn\\
&&+\frac12\{\hat F_{\a\r}\ds \hat D_\b\hat{\phi}\}
\star \hat D_\s\hat{\phi}\star \hat{\phi}+\frac12D_\r \hat{\phi}\star\{\hat 
F_{\a\s}\ds \hat
D_\b\hat{\phi}\}\star\hat{\phi}\Big)\Big)^{(1)} . 
\label{NCDejstvo_S_3_prvidrugi_red}
\end{eqnarray}
Explicit calculation then gives:
{\small
\begin{eqnarray}
S_{3NC}^{(2)} &=& -\frac{i}{256\pi
G_Nl^3}\theta^{\alpha\beta}\theta^{\gamma\delta}\varepsilon^{\mu\nu\rho\sigma}
\tr\int d^{4} x\,  \Bigg( \frac{1}{32}\{F_{\gamma\delta}, 
\{F_{\alpha\beta},D_{\mu}\phi
D_{\nu}\phi\}\} 
D_{\rho}\phi D_{\sigma} \phi \phi\nn\\
&&  -\frac18 \Big(\frac{i}{2}[D_{\gamma}F_{\alpha
\beta},D_{\delta}(D_{\mu}\phi
D_{\nu}\phi)]+\frac{1}{2}\{\{F_{\alpha\gamma},F_{\beta\delta}\},D_{\mu}\phi 
D_{\nu}\phi\}\nn\\
&& 
+\frac{i}{2}\{F_{\alpha\beta},(D_{\gamma}D_{\mu}\phi)(D_{\delta}D_{\nu}\phi)\} +
\frac{1}{2}\{F_{\alpha\beta}, [D_{\mu}\phi,\{F_{\gamma\nu},D_{\delta} 
\phi\}]\}\Big) D_{\rho}\phi
D_{\sigma} \phi \phi\nn\\
&&-\frac18\{F_{\alpha\beta},D_{\mu}\phi D_{\nu}\phi\}\Big(
\frac{i}{2}D_{\gamma}(D_{\rho}\phi D_{\sigma}\phi)D_{\delta}\phi\nn\\
&& +\frac{i}{2}(D_{\gamma}D_{\rho}\phi)(D_{\delta}D_{\sigma}\phi)\phi +
\frac{1}{2}[\{F_{\gamma\rho},D_{\delta}\phi\},D_{\sigma}\phi] \phi\Big)\nn\\
&&+\frac{i}{4}\Big(-\frac{1}{4}\{F_{\gamma\delta},(D_{\alpha}D_{\mu}\phi)(D_{
\beta} D_{\nu}\phi)\} 
D_{\rho}\phi D_{\sigma} \phi \phi +
\Big(\frac{i}{2}(D_{\gamma}D_{\alpha}D_{\mu}\phi)(D_{\delta}D_{\beta}D_{\nu}
\phi)\nonumber\\
&&+\frac{1}{2}\{\{F_{\gamma\alpha},D_{\delta}D_{\mu}\phi\},D_{\beta}D_{\nu}\phi
\} 
+\frac{1}{2}\{(D_{\alpha}\{F_{\gamma\mu},D_{\delta}\phi\}),(D_{\beta}D_{\nu}
\phi)\} \Big) D_{\rho}\phi D_{\sigma} \phi \phi\nonumber\\
&& + (D_{\alpha}D_{\mu} \phi)(D_{\beta}D_{\nu} \phi)\Big(
\frac{i}{2}D_{\gamma}(D_{\rho}\phi D_{\sigma}\phi)D_{\delta}\phi \nn\\
&& +\frac{i}{2}(D_{\gamma}D_{\rho}\phi)(D_{\delta}D_{\sigma}\phi)\phi +
\frac{1}{2}[\{F_{\gamma\rho},D_{\delta}\phi\},D_{\sigma}\phi] 
\phi\Big)\Big)\nn\\
&&+\frac{1}{4}\Big(-\frac{1}{4}\{F_{\gamma\delta},[\{F_{\alpha\mu},D_{\beta}
\phi\},D_{\nu}\phi]\} 
 D_{\rho}\phi D_{\sigma} \phi \phi + 
\Big(\frac{i}{2}\{D_{\gamma}\{F_{\alpha\mu},D_{\beta}\phi\},D_{\delta}D_{\nu}
\phi\}\nonumber\\
&&+\frac{i}{2}[[D_{\gamma}F_{\alpha\mu},D_{\delta}D_{\beta}\phi],D_{\nu}\phi] 
+\frac{1}{2}[\{\{F_{\alpha\gamma},F_{\mu\delta}\},D_{\beta}\phi\},D_{\nu}\phi]
\nn\\ 
&&+\frac{1}{2}[\{F_{\alpha\mu},\{F_{\gamma\beta},D_{\delta}\phi\}\},D_{\nu}\phi]
+\frac{1}{2}[\{F_{\alpha\mu},D_{\beta}\phi\},\{F_{\gamma\nu},D_{
\delta}\phi\}] \Big) D_{\rho}\phi D_{\sigma} \phi \phi \nn\\
&&+ \frac{i}{2}[\{F_{\alpha\mu},D_{\beta}\phi\},D_{\nu} \phi]\Big(
D_{\gamma}(D_{\rho}\phi D_{\sigma}\phi)D_{\delta}\phi
+(D_{\gamma}D_{\rho}\phi)(D_{\delta}D_{\sigma}\phi)\phi\nn\\ 
&& -i[\{F_{\gamma\rho},D_{ \delta}\phi\},D_{\sigma}\phi] \phi\Big) 
\Big) \nn\\
 &&+\frac{i}{4}\Big(-\frac{1}{4}\{F_{\gamma\delta},D_{\mu}\phi
D_{\nu}\phi\}[D_{\alpha}D_{\rho}\phi,D_{\sigma}\phi]D_{\beta}\phi
+ \Big(\frac{i}{2}(D_{\gamma}D_{\mu}\phi)(D_{\delta}D_{\nu}\phi)\nonumber\\
&&+\frac{1}{2}[\{F_{\gamma\mu},D_{\delta}\phi\},D_{\nu}\phi]\Big)[D_{\alpha}D_{
\rho}\phi,D_{\sigma}\phi]D_{\beta}\phi\nn\\
&& + D_{\mu}\phi D_{\nu} \phi\Big(
\frac{i}{2}D_{\gamma}([D_{\alpha}D_{\rho}\phi, 
D_{\sigma}\phi])D_{\delta}D_{\beta}\phi 
+\frac{i}{2}\{D_{\gamma}D_{\alpha}D_{\rho}\phi,D_{\delta}D_{\sigma}\phi\}D_{
\beta}\phi
\nonumber\\
&&+ \frac{1}{2}[\{F_{\gamma\alpha},D_{\delta}D_{\rho}\phi\},D_{\sigma}\phi] 
D_{\beta}\phi+
\frac{1}{2}[D_{\alpha}\{F_{\gamma \rho}, D_{\delta}\phi\}, 
D_{\sigma}\phi]D_{\beta}\phi\nn\\
&&+\frac{1}{2}[D_{\alpha}D_{\rho}\phi,\{F_{\gamma\sigma}, 
D_{\delta}\phi\}]D_{\beta}\phi
 +\frac{1}{2}[D_{\alpha}D_{\rho}\phi, D_{\sigma}\phi]\{F_{\gamma\beta}, 
D_{\delta}
\phi\}\Big)\Big) \nn\\
&&+\frac{i}{4} \Big(-\frac{1}{4}\{F_{\gamma\delta},D_{\mu}\phi D_{\nu}\phi\} 
 (D_{\alpha}D_{\rho}\phi) (D_{\beta}D_{\sigma} \phi) \phi + 
\Big(\frac{i}{2}(D_{\gamma}D_{\mu}\phi)(D_{\delta}D_{\nu}\phi)\nonumber\\
&&+\frac{1}{2}[\{F_{\gamma\mu},D_{\delta}\phi\},D_{\nu}\phi]\Big)(D_{\alpha}D_{
\rho}\phi)(D_{\beta}D_{\sigma}\phi)\phi  \nonumber\\
&& + D_{\mu} \phi D_{\nu} \phi\Big(
\frac{i}{2}D_{\gamma}((D_{\alpha}D_{\rho}\phi) 
(D_{\beta}D_{\sigma}\phi))D_{\delta}\phi 
+\frac{i}{2}(D_{\gamma}D_{\alpha}D_{\rho 
}\phi)(D_{\delta}D_{\beta}D_{\sigma}\phi)\phi \nn\\
&& + 
\frac{1}{2}\{\{F_{\gamma\alpha},D_{\delta}D_{\rho}\phi\},D_{\beta}D_{\sigma}
\phi\} \phi +
\frac{1}{2}\{(D_{\alpha}\{F_{\gamma\rho},D_{\delta}\phi\}),
D_{\beta}D_{\sigma}\phi\}\phi\Big)\Big)\nn\\
&& +\frac{i}{4}\Big(-\frac{1}{4}\{F_{\gamma\delta},[\{F_{\alpha\rho},
D_{\beta}\phi\},D_{\sigma}\phi]\}\phi D_{\mu}\phi D_{\nu}\phi
+ \Big(\frac{i}{2}\{D_{\gamma}\{F_{\alpha\rho}, 
D_{\beta}\phi\},D_{\delta}D_{\sigma}\phi\}\nonumber\\
&&\frac{i}{2}[[D_{\gamma}F_{\alpha\rho},D_{\delta}D_{\beta}\phi],D_{\sigma}\phi]
+\frac{1}{2}[\{\{F_{\alpha\gamma},F_{\rho\delta}\},D_{\beta}\phi\},D_{\sigma}
\phi]
\nonumber\\
&& 
+\frac{1}{2}[\{F_{\alpha\rho},\{F_{\gamma\beta},D_{\delta}\phi\}\},D_{\sigma}
\phi] 
+\frac{1}{2}[\{F_{\alpha\rho},D_{\beta}\phi\},\{F_{\gamma\sigma},D_{\delta} 
\phi\}]\Big)\phi D_{\mu}\phi D_{\nu}\phi \nn\\
&& + 
[\{F_{\alpha\rho},D_{\beta}\phi\},D_{\sigma}\phi]\Big(\frac{i}{2}D_{\gamma}\phi 
D_{\delta}(D_{\mu}\phi D_{\nu}\phi) \nn\\
&& +\frac{i}{2}\phi(D_{\gamma}D_{\mu}\phi)(D_{\delta}D_{\nu}\phi) + 
\frac{1}{2}\phi[\{F_{\gamma\mu},D_{\delta}\phi\},D_{\nu}\phi]\Big)\Big)
\Bigg)  .\label{NCDejstvo_S_3_drugi_red}
\end{eqnarray}
}
The expanded actions (\ref{NCDejstvo1Exp2}), (\ref{NCDejstvo_S2-drugi-red-2}) 
and (\ref{NCDejstvo_S_3_drugi_red}) are obviously invariant under the 
commutative $SO(2,3)$
gauge transformations, as guaranteed by the SW map. 

\initiate
\section{Symmetry breaking and the low energy expansion}

\noindent The second order expansion of the NC actions (\ref{NCdejstvo_S_1},
\ref{NCDejstvo_S_2}, \ref{NCDejstvo_S_3}), given by equations
(\ref{NCDejstvo1Exp2}, \ref{NCDejstvo_S2-drugi-red-2}, 
\ref{NCDejstvo_S_3_drugi_red}) is
explicitly invariant under the commutative $SO(2,3)$ gauge symmetry. In order 
to 
relate these
expanded actions to the General Relativity and its NC corrections, we have to 
follow the same steps
as in the commutative model, Section 2. First we have to break the $SO(2,3)$ 
gauge
symmetry down to the $SO(1,3)$ gauge symmetry (local Lorentz symmetry). Then we 
have to calculate
the traces and write the actions in terms of the geometric quantities 
(curvature, vierbeins,
metric). Let us proceed step by step. 

\noindent The symmetry breaking is done by choosing the field $\phi$ to be of 
the form
$\phi=(0,0,0,0,l)$. In this way the zeroth order of the actions 
(\ref{NCdejstvo_S_1},
\ref{NCDejstvo_S_2}, \ref{NCDejstvo_S_3}) reduces to the commutative model with 
the BG term,
Einstein-Hilbert term and the cosmological constant term, 
(\ref{FullCommAction}). Then we have to
calculate the
traces. As we have seen, the first order correction vanishes and the second 
order correction is the first non-vanishing correction. It is very long and we 
will not
write the full expressions here. Moreover, the expended actions contain 
terms that are
fourth and lower
powers of curvature and second and lower powers of torsion. To analyze the full 
action is very demanding. Especially, 
finding equations of
motion is a highly non trivial calculation. Additionally, there is no guarantee 
that the obtained equations of motion
will remain second order partial differential equations with respect to the 
metric and the
connection. There are higher order gravity theories, like Lovelock theories, 
where the equations of
motion remain the second order differential equations. In our case, 
unfortunately  it is not clear
what will happen  with the equations of motion and a careful analysis has to 
be done. 

\subsection{Low energy effective NC gravity action}

\noindent However, we can still analyze different sectors of our model, such as 
high energy
behavior, or low energy behavior, with or without the cosmological 
constant, etc. In
this paper we are interested in the low energy corrections. To be more precise, 
we keep terms that have at most two derivatives
on vierbeins. Therefore, in our analysis we include terms linear in curvature, 
linear and quadratic in 
torsion. Additionally, we assume that the spin connection $\omega_\mu^{ab}$ and 
first order derivatives of vierbeins such as $\partial_\rho e_\a^b$ are of the 
same order. 

The low energy NC correction of the action $S_{1NC}$ is given by
\begin{eqnarray} 
S_{1NC}^{(2)} &=&-\frac{1}{128\pi G_Nl^4}\int {\rm d}^{4} x\,  e
\theta^{\alpha\beta}\theta^{\gamma\delta}\Big(
2R_{\alpha\beta\gamma\delta} 
-4R_{\alpha\gamma\beta\delta}+6g_{\beta\delta}{R_{\alpha\mu\gamma}}^{\mu} 
\nn\\&&
-\frac{6}{l^{2}} g_{\alpha\gamma} g_{\beta\delta} 
-5 T_{\alpha\beta}^a T_{\gamma\delta 
a} +10T_{\alpha\gamma}^a 
T_{\beta\delta a}
-3T_{\a\b\g}T_{\d\m}^{\ \ \m}- T_{\a\b\r}T^\r_{\ 
\g\d}
-8T_{\a\g\d}T_{\b\m}^{\ \ \m}\nn\\&&
-2T_{\alpha\g\m}e_\b^b\nabla_\delta e_\m^b  
-2T_{\alpha\g\beta}e^\r_a\nabla_\delta 
e_\rho^a+6T_{\d\r\beta}e^\r_a\nabla_\a e_\g^a
\nn\\
&&-2T_{\a\beta\d}e^\r_a\nabla_\g e_\r^a
+T_{\a\beta}^{\ \ \m}e_{\d a}\nabla_\g e_\m^a
  +4e^{\m}_ae_{b\b}
\nabla_{\g}e_{\a}^{a}\nabla_{\d}e_{\m}^{b}+4e^{\m}_ae_{b\d}
\nabla_{\g}e_{\b}^{b}\nabla_{\a}e_{\m}^{a}
\nn\\
&&+2g_{\a\g}e^{\m}_ae^{\n}_b
\nabla_{\b}e_{\m}^{a}\nabla_{\d}e_{\n}^{b}
-2g_{\a\g} e^\m_b e^\n_a\nabla_\d e_\n^b \nabla_\b e_{\m }^a
\Big) .\nn 
\end{eqnarray}
The low energy NC correction of the action $S_{2NC}$ is given by
\begin{eqnarray} 
S_{2NC}^{(2)} &=&\frac{1}{256\pi G_Nl^4}\int {\rm d}^{4} x\,  e
\theta^{\alpha\beta}\theta^{\gamma\delta}\Big(
20R_{\alpha\beta\gamma\delta} 
-28R_{\alpha\gamma\beta\delta}-56g_{\beta\delta}{R_{\alpha\mu\gamma}}^{\mu} 
\nn\\&&
+\frac{68}{l^{2}} g_{\alpha\gamma} g_{\beta\delta}+
T_{\alpha\beta}^a 
T_{\gamma\delta 
a} -11T_{\alpha\gamma}^a 
T_{\beta\delta a}
+6T_{\a\b\r}T_{\r\g\d}-16T_{\a\g\b}T_{\d\m}^{\ \ \m}\nn\\&&
+24 T_{\a\g}^{\ \ \r}T_{\r\d\b}+4g_{\b\d}T_{\g\s}^{\ \ \s} T_{\a\r}^{\ \ 
\r}-4g_{\b\d}T_{\g\s\r}T_{\a}^{\ \r\s}\nn\\
&&+4T_{\alpha\g\d}e^\m_b\nabla_\b e_\m^b +4T_{\alpha\b\g}e^\m_b\nabla_\d 
e_\m^b+2T_{\alpha\beta\r}e_\g^a\nabla_\delta 
e_\rho^a\nn\\
&&-12T_{\a\g}^{\ \ \r} e_{\d a}\nabla_\b e_\r^a
-28T_{\a\m\g}e^\m_a\nabla_\b e_\d^a
+4g_{\b\g}T_{\a\r}^{\ \ \s}e^\r_b\nabla_\d e_\s^b
-4g_{\a\g}T_{\n\beta}^{\ \ \n}e_{\m}^a\nabla_\d e_\m^a\nn\\
&&-40
e^{\r}_ae_{\d b}
\nabla_{\a}e_{\g}^{a}\nabla_{\b}e_{\r}^{b}
-12g_{\b\d} e^\m_b e^\n_a\nabla_\g 
e_\m^b \nabla_\a e_{\n }^a\nn\\&&+32 e^{\m}_be_{\d a}
\nabla_{\a}e_{\g}^{a}\nabla_{\b}e_{\m}^{b}
+12 g_{\b\d}e^\r_c e^\m_a \nabla_\a e_\r^a \nabla_\g e_\m^c \Big) .\nn 
\end{eqnarray}
The low energy NC correction of the action $S_{3NC}$ is given by 
\bea
S_{3NC}^{(2)}&&=
\int d^{4} x\, 
\frac{  e 
\theta^{\alpha\beta}\theta^{\gamma\delta}}{128\pi G_N l^4}
 \Big( 38 
R_{\alpha\beta\g\delta}-44R_{\alpha\gamma\beta\delta}-36R_{\alpha\gamma}g_{
\beta\delta}+\frac{56}{l^{2}} 
g_{\alpha\gamma} g_{\beta\delta}\nn\\
&&-7 T_{\alpha\beta}^{a} T_{\gamma\delta a}
+14T_{\alpha\gamma}^{a} T_{\beta\delta 
a}-2T_{\alpha\beta\gamma}T_{\delta\rho}^{\rho}+ 
4T_{\alpha\gamma}^{\rho}T_{\delta\rho\beta}
+4 g_{\beta\delta}T_{\alpha\rho}^{\rho}T_{\gamma\sigma}^{\sigma}\nn\\
&&-4g_{\beta\delta} T_{\alpha\rho}^{\sigma}T_{\gamma\sigma}^{\rho}+
32\nabla_{\alpha}e_{\gamma}^{a}e_{\delta 
a}\nabla_{\beta}e_{\rho}^{b}e^{\rho}_{b} 
-32\nabla_{\alpha}e_{\gamma}^{a}e^{\rho}_{a}\nabla_{\beta}e_{\rho}^{b}e_{\delta 
b}\nn\\&&
+8g_{\beta\delta}\nabla_{\alpha} e_{\rho}^a e^{\sigma}_a\nabla_{\gamma} 
e_{\sigma}^b e^{\rho}_b
-8g_{\beta\delta}\nabla_{\alpha} e_{\rho}^a e^{\rho}_a\nabla_{\gamma} 
e_{\sigma}^b e^{\sigma}_b
-12T_{\alpha\gamma}^{\rho}\nabla_{\beta} e_{\rho}^a e_{\delta a}\nn\\
&&+18T_{\alpha\beta\gamma}\nabla_{\delta}e_{\rho}^{a} 
e^{\rho}_{a}-8T_{\alpha\gamma\beta}\nabla_{\delta}e_{\rho}^{a} e^{\rho}_{a} 
+16T_{\gamma\rho\beta}\nabla_{\alpha}e_{\delta}^{a} e^{\rho}_{a} \nn\\
&&-4T_{\alpha\rho}^{\sigma}\nabla_{\gamma}e_{\sigma}^{a} 
e^{\rho}_{a}g_{\beta\delta} +
4T_{\alpha\rho}^{\rho}\nabla_{\gamma}e_{\sigma}^{a} 
e^{\sigma}_{a}g_{\beta\delta}\Big) .
\eea
Remembering that $c_1+c_2=1$ the resulting action follows
\begin{eqnarray} 
S_{NC} &=&
-\frac{1}{16\pi G_{N}}\int 
{\rm d}^{4}x\, e\Big(R-\frac{6}{l^2}(1+c_2+2c_3)\Big)\nn\\
&& +\frac{1}{128\pi G_Nl^4}\int {\rm d}^{4} x\,  e
\theta^{\alpha\beta}\theta^{\gamma\delta}\Big(
(-2+12c_2+38c_3)R_{\alpha\beta\gamma\delta} \nn\\
&&+(4-18c_2-44c_3)R_{\alpha\gamma\beta\delta} 
-(6+22c_2+36c_3)g_{\beta\delta}{R_{\alpha\mu\gamma}}^{\mu} \nn\\
&&+\frac{6+28c_2+56c_3}{l^{2}} g_{\alpha\gamma} g_{\beta\delta} 
+(5-\frac92c_2-7c_3) T_{\alpha\beta}^a T_{\gamma\delta 
a}\label{action-linearnoRT}\\
&& +(-10+\frac92c_2+14c_3)T_{\alpha\gamma}^a 
T_{\beta\delta a}
+(3-3c_2-2c_3)T_{\a\b\g}T_{\d\m}^{\ \ \m}\nn\\&&+(1+2c_2) T_{\a\b\r}T^\r_{\ 
\g\d}
+8T_{\a\g\d}T_{\b\m}^{\ \ \m}\nn\\&&-(2c_2+4c_3) T_{\a\g\r}T^\r_{\ 
\d\b}
+(2c_2+4c_3)g_{\b\d}T_{\g\s}^{\ \ \s}T_{\a\r}^{\ \ \r}\nn\\
&& -(2c_2+4c_3)T_{\a\r\s}T_\g^{\ \s\r}g_{\beta\delta}
+(-2+4c_2+18c_3)T_{\alpha\beta\g}e^\r_a\nabla_\delta e_\rho^a \nn\\ 
&&+(6-8c_2-8c_3)T_{\alpha\g\beta}e^\r_a\nabla_\delta e_\rho^a
+(2+4c_2+12c_3)T_{\alpha\g}^{\ \ \m}e_\b^a\nabla_\delta e_\m^a\nn\\
&&-T_{\alpha\beta}^{\ \ \m}e_\d^a\nabla_\gamma e_\m^a
+
(-6-8c_2-16c_3)T_{\d\r\beta}e^\r_a\nabla_\a e_\g^a
\nn\\
&&-(2c_2+4c_3)g_{\a\g}T_{\m\b}^{\ \ \m}e^\r_a\nabla_\d e_\r^a
-(2c_2+4c_3)g_{\b\d}T_{\a\r}^{\ \ \s} e_a^\rho \nabla_\g e_\s^a 
\nn\\   
&&-(4+16c_2+32c_3)e^{\m}_ae_{b\b}
\nabla_{\g}e_{\a}^{a}\nabla_{\d}e_{\m}^{b}
+(4+12c_2+32c_3)e_{\d a}e_{b}^\m
\nabla_{\a}e_{\g}^{a}\nabla_{\b}e_{\m}^{b}\nn\\
&&-(2+4c_2+8c_3)g_{\b\d}e^{\m}_ae^{\n}_b
\nabla_{\g}e_{\m}^{a}\nabla_{\a}e_{\n}^{b}
+(2+4c_2+8c_3)g_{\b\d} e^\m_a e^\r_c\nabla_\a e_\r^a \nabla_\g e_{\m }^c \Big)
.\nn 
\end{eqnarray}
To obtain this action we used that $D_\a F_{\m\n}$ is the $SO(2,3)$ covariant 
derivative and its components are
\bea
&&(D_\a F_{\m\n})^{ab} = \nabla_\a F_{\m\n}^{\ \ ab} - \frac{1}{l^2}(e_\a
^aT_{\m\n}^{\ b} -e_\a ^bT_{\m\n}^{\ a}) ,\nn\\
&&(D_\a F_{\m\n})^{a5} = \frac{1}{l}( \nabla_\a T_{\m\n}^{a} + e_\a^m
F_{\m\n m}^{\ \ \ \ a}) \, ,\nn\\
&&(D_\kappa D_\a F_{\m\n})^{ab} = \nabla_\kappa\nabla_\a F_{\m\n}^{\ \ ab} -
\frac{1}{l^2}\Big( 
(\nabla_\kappa e_\a ^a)T_{\m\n}^{\ b} -(\nabla_\kappa e_\a ^b)T_{\m\n}^{\ a} +
e_\a ^a(\nabla_\kappa T_{\m\n}^{\ b})\nn\\
&& - e_\a ^b(\nabla_\kappa T_{\m\n}^{\ a})
+ e_\kappa ^a(\nabla_\a T_{\m\n}^{\ b}) - e_\kappa ^b(\nabla_\a T_{\m\n}^{\ a}) 
+ e_\kappa ^a e^m_\a F_{\m\n m}^{\ \ \ \ b} - e_\kappa ^b e^m_\a F_{\m\n m}^{\ \
\ \ a} \Big) \ ,\nn
\eea
with the $SO(1,3)$ covariant derivative 
\begin{eqnarray}
\nabla_\a F_{\m\n}^{\ \ ab} &=&
\partial_\a F_{\m\n}^{\ \ ab} + \omega_\alpha^{ac}F_{\mu\nu c}^{\ \ \ b}
-\omega_\alpha^{bc}F_{\mu\nu c}^{\ \ \ a},\nn\\ 
\nabla_\a T_{\m\n}^{a} &=& \partial_\a T_{\m\n}^{a} + \omega_\alpha^{ac} 
T_{\mu\nu c}.\nn
\end{eqnarray}
We also used that
\begin{eqnarray}
(D_\a \phi) ^{a} &=& e_\a^{\ a},\nn\\
(D_\a \phi) ^{5} &=&0 ,\nn\\ 
(D_\a D_\b \phi)^{a} &=&(\nabla_\a e_\b)^a ,\nn\\
(D_\a D_\b \phi)^{5} &=&-\frac{1}{l}g_{\a\b}\ .\nn 
\end{eqnarray}

\noindent Before we determine the equations of motion, let us briefly discuss 
the action (\ref{action-linearnoRT}). We see that
this action is invariant under the $SO(1,3)$ gauge 
symmetry. However, due to
the noncommutativity it
is no longer invariant under the diffeomorphism symmetry. The non-invariant 
terms manifest
themselves in two
ways. Firstly, there are tensors contracted with the NC parameter 
$\theta^{\a\b}$ such as
$\theta^{\a\b}\theta^{\k\l}R_{\alpha\kappa\beta\lambda}$. Since $\theta^{\a\b}$ 
is not a tensor
under the diffeomorphism symmetry (it is a constant matrix that does not 
transform under the
diffeomorphism), those terms are also
not scalars (tensors). Then there are terms in which $SO(1,3)$ covariant 
derivatives of vierbeins
appear. Using the metricity condition 
\begin{equation}
\nabla_\mu^{tot} e_\rho^{\ a} = \partial_\mu e_\rho^{\ a} + \omega_\mu^{ab} 
e_{\rho b} -
\Gamma_{\mu\rho}^\sigma e_\sigma^{\ a} =0 
\end{equation}
the $SO(1,3)$ covariant derivative can be written as
\begin{equation}
\nabla_\mu e_\rho^{\ a} = \partial_\mu e_\rho^{\ a} + \omega_\mu^{ab} e_{\rho b} 
=
\Gamma_{\mu\rho}^\sigma e_\sigma^{\ a} \ . \label{LorKovIzvTetrada}
\end{equation}
Therefore, the affine connection $\Gamma_{\mu\rho}^\sigma$ appears explicitly
in (\ref{action-linearnoRT}). Note that this affine connection does not have to 
be given
by the Christoffel symbols. We will see in Section 6 that the noncommutativity
can generate the antisymetric part of the connection, leading to the appearance 
of torsion. 
Some of the terms with the explicit $\Gamma_{\mu\rho}^\sigma$s can be grouped to 
form the curvature
tensor, but some will remain and make
the diffeomorphism non-invariance explicit.

\subsection{Low energy equations of motion}

\noindent The equations of motions are obtained by varying the action 
(\ref{action-linearnoRT}) with respect to the vierbein and the spin 
connection. Some useful formulae are given in Appendix C.  
In this article we are interested in NC corrections to the GR solutions with 
vanishing torsion. Therefore, in the equations of motion we impose the 
condition $T_{\m\n}^a=0$. A more general form of the 
equations of motion will be presented in future work. 

\noindent Finally, the equation of 
motion for the 
vierbein is given by
\begin{equation}
 R_{\a\g}^{\ \ cd}e^\g_d e_a^\a e_c^\m-\frac12e^\m_a 
R+\frac{3}{l^2}(1+c_2+2c_3)e^\m_a=\tau_a^{\ \m} \, ,
\end{equation}
where 
\bea
\tau_a^{\ \m}&=&- \frac{8\pi G_N}{e} \frac{\delta S^{(2)}_{NC}}{\delta 
e_\mu^a}\nn\\
&=&-\frac{\theta^{\alpha\beta}\theta^{\gamma\delta}}{16l^4}\Big(
(-4+6c_2+22c_3)e^\m_a R_{\a\b\g\d}\nn\\
&&+(4-18c_2+44c_3)(e^\m_a R_{\a\g\b\d}-
2\d_\a^\m R_{\b\d \g a})
\nn\\
&&-(6+22c_2+36c_3)e^\m_a g_{\b\d}R_{\a\g}
+2\d_\a^\m e_{\g a}R_{\b\d}+g_{\b\d}R_{\g \l}e^{\l a} \d_\a^\m
-g_{\b\d}R_{\a\l\g}^{\ \ \ \ 
\m}e_a^\l)
\nn\\
&&+(4+16c_2+32c_3)(e^\m_c e_{\g b}e^\rho_a\nabla_\b e_\d^c\nabla_\a 
e_\rho^b\nn\\ 
&&- e^\m_a e_{\b b}e^\rho_c\nabla_\g e_\a^c\nabla_\d e_\rho 
^b - \d^\m_\a e^\rho_b\nabla_\g e_{\r a}\nabla_\d e_\b^b)\nn\\
&&+(2+2c_2+4c_3)g_{\b\d}e_a^\m  e^\rho_c e_d^\s( \nabla_\a e_{\r }^c\nabla_\g 
e_\s^d- \nabla_\a e_{\r }^d\nabla_\g e_\s ^c)\nn\\
&&+(4+6c_2+12c_3)(e^\m_ag_{\b\d}e_b^\r\nabla_\a \nabla_\g e_\r ^b -e_{\d b} 
e_c^\rho e^\m_a\nabla_\b e_{\r }^c\nabla_\g e_\a ^b \nn\\
&&+g_{\b\d}e_a^\r e^\s_b e^\m_c\nabla_\a e_\s^c \nabla_\g e_\r 
^b-g_{\b\d}e_a^\r e^\m_b\nabla_\a \nabla_\g e_\r ^b)\nn\\
&&-(4+12c_2+32c_3)e^\m_a\nabla_\b e_{\d b}\nabla_\g 
e_{\a}^b+(7-14c_2-54c_3)\d^\m_\a R_{\g\d\b a}\nn\\
&&+(5+12c_2+24c_3)(2e_{\g a} e_b^\m e_c^\s
\nabla_\d e_\b^b\nabla_\a e_{\s }^c + 2 e_b^\m 
\nabla_\a e_{\g a}\nabla_\d e_{\b }^b \nn\\
&&- 2e_{\g a} e_b^\s  e^\m_c
\nabla_\a e_\s^c\nabla_\d e_{\b }^b-e_{\d a}R_{\a\b\ \g}^{\ \ \ \m})\nn\\
&&+(2+8c_2-12c_3)\d^\m_\a e_{\d a}( e_c^\s e_b^\r
\nabla_\b e_\s^c\nabla_\g e_{\r }^b- e_b^\s e_c^\r
\nabla_\b e_\s^c\nabla_\g e_{\r }^b)\nn\\
&&+(6+24c_2+36c_3)\d^{\m}_{ \a} e_b^\r
\nabla_\g e_\r^b\nabla_\b e_{\d a}+2(-2+4c_2+18c_3)\d_\a^\m e_{\g a} 
e_b^\r\nabla_\b \nabla_\d e_\r ^b\nn\\
&&+(2+4c_2+12c_3)\d_\a^\m e_{\g b} 
e_a^\r\nabla_\d \nabla_\b e_\r ^b
\nn\\
&&-(6+8c_2+16c_3)\d_\a^\m(e_{\g a}  e_c^\s e_b^\r
\nabla_\d e_\b^b\nabla_\r e_{\s }^c+  e_d^\r
\nabla_\d e_\b^d\nabla_\r e_{\g a}\nn\\
&&- e_{\g a}  e_c^\r
e^\s_d\nabla_\r e_\s^c\nabla_\d e_{\b }^d + e_{\g a}  e_d^\r
\nabla_\r \nabla_\d e_{\b }^d)\nn\\
&&+(6-8c_2-8c_3)\d^\m_\a e_{\g a}  e_d^\r
\nabla_\d\nabla_\b e_{\r }^d\nn\\
&&+(2+2c_2+8c_3)\d_\a^\m  e_{\d b}(e^\r_c  e_a^\s 
\nabla_\g e_\s^c\nabla_\r e_{\b }^b-e^\s_c  e_a^\r 
\nabla_\g e_\s^c\nabla_\b e_{\r }^b)\nn\\
&&+(6+22c_2+48c_3)\d_\a^\m e_{\b b}(e^\s_c  e_a^\r 
\nabla_\g e_\s^c\nabla_\d e_{\r }^b-e^\s_a  e_c^\r 
\nabla_\g e_\s^c\nabla_\d e_{\r }^b)\nn\\
&&-2\d_\a^\m(e^\s_c  e_a^\r e_{\d b}
\nabla_\b e_\s^c\nabla_\g e_{\r }^b-e^\r_c  e_a^\s e_{\d b}
\nabla_\b e_\s^c\nabla_\g e_{\r }^b-e^\r_a  e_\g^b 
\nabla_\b \nabla_\d e_{\r }^b )\nn\\
&&-(8+20c_2+44c_3)\d_\a^\m  e_a^\r 
\nabla_\g e_\r^b\nabla_\b e_{\d }^b\nn\\
&&+(2c_2+4c_3)g_{\b\d}(\d_\a^\m  e_b^\r e^\s_a
\nabla_\s \nabla_\g e_{\r }^b\nn\\
&&-\d_\a^\m  e_c^\s e^\r_a
\nabla_\s \nabla_\g e_{\r }^c - \d_\a^\m  e^\n_a e_c^\r e^\s_d
\nabla_\n e_\d^c \nabla_\g e_{\r }^d\nn\\
&&-\d_\a^\m  e^\n_d e_c^\r e^\s_a
\nabla_\r e_\n^d \nabla_\g e_{\s }^c-  e^\r_a e_b^\n e^\s_c
\nabla_\n e_\r^c \nabla_\g e_{\s }^b
- e^\s_a e_c^\n e^\r_d
\nabla_\r e_\n^d \nabla_\g e_{\s }^c)\nn\\
&&+\frac{6+28c_2+56c_3}{l^2}(g_{\a\g}g_{\b\d} e^{\m}_a+4g_{\b\d}\d_\a^\m 
e_{\g}^a)
\Big)\ .
\eea
Multiplying the previous equation with $e_a^\n$ and using the metricity 
condition we obtain
\be 
R^{\n\m}-\frac12g^{\m\n}R+\frac{3}{l^2}(1+c_2+2c_3) g^{\m\n}=\tau^{\m\n} 
,\label{EoM-metric-0}
\ee
with
\bea
\tau^{\m\n} &=&-\frac{\theta^{\a\b}\theta^{\g\d}}{16l^4}\Big(
(-4+6c_2+22c_3)R_{\a\b\g\d}g^{\m\n}-(7-14c_2-54c_3)R^\n_{\ 
\b\g\d}\d_\a^\m\nn\\
&& +(4-18c_2-44c_3)(2R^\n_{\ 
\g\b\d}\d_\a^\m+R_{\a\g\b\d}g^{\m\n})\nn\\
&&-(5+12c_2+24c_3)R^\m_{\ \g\a\b}\d^\n_\d\nn\\
&&-(6+22c_2+36c_3)(g^{\m\n}
g_{\b\d}R_{\a\g}+2\d_\b^\m \d_\d^\n R_{\a\g}+g_{\b\d}\d_\g^\m R_{\a}^{\ 
\n}-g_{\b\d}R_{\a\ \g}^{\ \n\ \m})\nn\\ 
&&+(4+16c_2+32c_3)(g_{\s\b}g^{\r\n}
\G^{\m}_{\a\g}\G^{\s}_{\d\r}-g^{\m\n} g_{\r\b}\G^{\s}_{\a\g}\G^{\r}_{\s\d})\nn\\
&&+(2+2c_2+4c_3)g^{\m\n} 
g_{\d\b}\G^{\s}_{\a\s}\G^{\r}_{\g\r}-(4+12c_2+32c_3)g^{\m\n} 
g_{\r\s}\G^{\r}_{\b\d}\G^{\s}_{\g\a}\nn\\
&&+(2+4c_2+8c_3)g^{\m\n}g_{\b\d}\G^\s_{\g\r}\G^\r_{\a\s}\nn\\
&&+(4+6c_2+12c_3)(g^{\m\n}(g_{\b\d}\pa_\a\G^\r_{\g\r}-g_{\d\r}\G^\s_{\b\s}\G^\r_
{ \a\g})\nn\\
&&+g^{\r\n}(g_{\a\s} \G^\s_{\b\d}\G^\m_{\g\r}+g_{\a\d} 
\pa_\b\G^\m_{\g\r}))\nn\\
&&+(10+24c_2+48c_3)(
\d^\n_\g\G^\r_{\a\r}\G^\m_{\b\d}+
\G^\m_{\d\b}\G^\n_{\a\g}-\d^\n_\g\G^\m_{\a\s}\G^\s_{\b\d})\nn\\
&&- (4+4c_2+8c_3)g_{\b\d}g^{\s\n}
\G^{\r}_{\g\r}\G^{\m}_{\a\s}+(6+24c_2+36c_3)\d_\a^\m\G^{\r}_{\g\r}\G^{\n}_{
\b\d }\nn\\
&&+(2+8c_2-12c_3)\d^\m_\a 
\d^\n_\d(\G^\s_{\b\s}\G^\r_{\g\r}-\G^\r_{\b\s}\G^\s_{\g\r})\nn\\
&&+(-4+8c_2+36c_3)
\d_\g^\n\d_\a^\m\pa_\b\G^\r_{\d\r}+(6-8c_2-8c_3)
\d_\g^\n\d_\a^\m\pa_\d\G^\r_{\d\b}\nn\\
&&+(2+28c_3)\d^\m_\a 
\d^\n_\g \G^\s_{\d\r}\G^\r_{\b\s}+2\d_\a^\m 
g^{\r\n}g_{\g\kappa}(\pa_\b\G^\kappa_{\d\r}+\G^\s_{\d\r}\G^\kappa_{\b\s})\nn\\
&&-(6+8c_2+16c_3)\d_\a^\m(\d^\n_\g\G^{\s}_{\r\s}\G^{\r}_{
\b\d }+\G^{\n}_{\r\g}\G^{\r}_{
\b\d }+\d_\g^\n\pa_\r\G^{\r}_{\b\d})\nn\\
&&+(2+2c_2+8c_3)\d^\m_\a g^{\r\n}g_{\tau\d}(-\G^\s_{\g\s}\G^\tau_{
\b\r } + 
\G^\s_{\g\r}\G^\tau_{\b\s})\nn\\
&&+(6+22c_2+48c_3)\d^\m_\a g^{\r\n}g_{\tau\b}(\G^\s_{\g\s}\G^\tau_{
\d\r } - 
\G^\s_{\g\r}\G^\tau_{\d\s})\nn\\
&&-2\d_\a^\m g_{\kappa\d}g^{\r\n}(\G^\tau_{\b\tau}\G^\kappa_{
\g\r }-\G^\s_{\b\r}\G^\kappa_{
\g\s })\nn\\
&&+(2+4c_2+12c_3)
\d_\a^\m g_{\tau\g}g^{\r\n}(\pa_\d\G^\tau_{
\b\r }
+\G^\s_{\b\r}\G^\tau_{\d\s})\nn\\
&&-(8+20c_2+44c_3)
\d_\a^\m g_{\tau\s}g^{\r\n}\G^\s_{\g\r}\G^\tau_{\d\b}\nn\\
&&-(4+16c_2+32c_3)
\d_\a^\m 
\G^\r_{\b\d}\G^\n_{\g\r}-(2c_2+4c_3)g_{\b\d}g^{\s\n}\G^\r_{\a\s}\G^\m_{\g\r}
\nn\\
&&+(2c_2+4c_3)
\d_\a^\m g_{\b\d}g^{\r\n}(\pa_\r\G^\s_{
\g\s }-\pa_\s\G^\s_{
\g\r } 
-\G^\s_{\g\r}\G^\tau_{\tau\s}+\G^\s_{\r\kappa}\G^\kappa_{\g\s})
\nn\\
&&+\frac{
6+28c_2+56c_3}{l^2}(g^{\m\n}g_{\a\g}g_{\b\d}+4g_{\b\d}
\d_\g^\n\d_\a^\m)\Big)\ .\label{EoM-metric}
\eea
The equation of motion obtained by varying the action 
(\ref{action-linearnoRT}) with respect to the spin connection is given by
\begin{equation} 
T_{ac}^{\ \ c}e_b^\m-T_{bc}^{\ \ c}e_a^\m-T_{ab}^{\ \ \m}=S_{ab}^{\ \ \m}\ ,
\end{equation}
where 
\bea
S_{ab}^{\ \ \m}&=& -\frac{16\pi G_N}{e}\frac{\delta S^{(2)}_{NC}}{\delta 
\omega_\mu^{ab}}\nn\\
&=&-\frac{\theta^{\a\b}\theta^{\g\d}}{8l^4}\Big(
(2-4c_2-36c_3)\d_\a^\m\G_{\b\r}^\r e_{\g b}e_{\d a}\nn\\
&& -(1+10c_2+22c_3)\d_\a^\m\G_{\g\r}^\r(e_{\b a}e_{\d b}-e_{\b b}e_{\d a})\nn\\
&& - (5+6c_2-8c_3)\d_\a^\m\G_{\g\b}^\s(e_{\s a}e_{\d b}-e_{\s b}e_{\d a})\nn\\
&& - (3+12c_2+20c_3)g_{\b\d}\Big(\G_{\a\r}^\r(e_{\g b}e_{ a}^\m-e_{\g a}e_{ 
b}^\m)-\G_{\a\r}^\m(e_{\g b}e^\r_a-e_{\g a}e^\r_b)\Big)\nn\\
&& -(3+11c_2+18c_3)\Big(g_{\b\d}\G_{
\a\g}^\s(e_{\s b}e^\m_a-e_{\s a}e^\m_b)
+g_{\b\s}\G_{\a\d}^\s(e_{\g b}e^\m_a-e_{\g 
a}e^\m_b)\Big)\nn\\
&& - (5+14c_2+24c_3)\d_\a^\m g_{\d\b}\G_{\s\g}^\r  (e_{\r 
a}e^\s_{ 
b}-e_{\r b}e^\s_{ a})-g_{\d\s}\d^\m_\a\G_{\g\n}^\s (e_{\b 
b}e^\n_{a}-e_{\b a}e^\n_{ b})\nn\\
&& +(c_2-4c_3)\d_\a^\m g_{\d\s}\G_{\b\n}^\s(e_{\g 
b}e^\n_a-e_{\g a}e^\n_b)\nn\\
&& +(4+13c_2+24c_3)\d_\a^\m g_{\s\b}\G_{\d\n}^\s(e_{\g 
b}e^\n_a-e_{\g a}e^\n_b)
\Big)\ .\label{EoM-torsion}
\eea
The equations (\ref{EoM-metric}) and (\ref{EoM-torsion}) have a very clear 
physics 
interpretation. The noncommutativity is a source curvature and torsion, i.e. 
flat space-time 
becomes curved  as an effect of noncommutative
corrections. Also, a torsion-free solution will develop a non-zero torsion in 
the presence of
noncommutativity. 

\initiate
\section{NC Minkowski space-time}

In order to investigate consequences of noncommutativity in more details we 
analyze the NC deformation of 
Minkovski space-time. Minkowski space-time is a vacuum solution of the Einstein 
equations without
the cosmological constant. Therefore, we first have to assume that 
$1+c_2+2c_3 =0$, that is that the cosmological constant is not present in the 
zeroth order in the deformation parameter. Note that in our previous work 
\cite{MiAdSGrav, MDVR-14} we were not able to choose
the value of the cosmological constant, since we only worked with the action 
$S_{1NC}$. Adding the
other two actions $S_{2NC}$ and $S_{3NC}$ with arbitrary constants 
$c_1,c_2,c_3$ enables us to
study a wider class of NC gravity solutions. Assuming that the solution is a 
small perturbation around the flat Minkowski metric
\be 
g_{\m\n}=\eta_{\m\n}+h_{\m\n} ,
\ee
where $h_{\m\n}$ is a small correction of the second order in the deformation 
parameter
$\theta^{\mu\nu}$, equation (\ref{EoM-metric}) reduces to 
\be  
\frac{1}{2}(\pa_\s\pa^\n h^{\s\m}+\pa_\s\pa^\m h^{\s\n}-\pa^\m\pa^\n h -\Box 
h^{\m\n})-\frac12\eta^{\m\n}(\pa_\a\pa_\b h^{\a\b}-\Box h)=\tau^{\m\n} , 
\label{NCMinkEoMmetric} 
\ee
with 
\be 
\tau^{\m\n}=\frac{11}{4l^6}(2\eta_{\a\g}\theta^{\a\m}\theta^{\g\n}+\frac{1}{2}g_
{\a\g}
g_{\b\d} g^{\m\n}\theta^{\a\b}\theta^{\g\d}) . \nn
\ee
The second equation (\ref{EoM-torsion}) gives no contribution, that is the NC 
Minkowski space-time
remains torsion-free in the second order of the deformation parameter. The 
small perturbation
$h_{\mu\nu}$ we split into components $h_{00}$, $h_{0j}$ and $h_{ij}$ and 
we write equations separately for each component. Note
that $i,j,\dots$ are space indices, they take values $1,2,3$ and we label 
$\psi=\d_{ij}h^{ij}$. The $00,0j$ and $ij$ components of (\ref{NCMinkEoMmetric}) 
are
given by: 
\bea
&&\triangle\psi-\pa_i\pa_j h^{ij}=2\tau^{00} ,\label{00}\\
 &&\pa_0\pa_j h^{ij}-\pa_i\pa_j h^{j0}-\pa_0\pa_i \psi+\triangle 
h^{0i} =2\tau^{0i},\label{0j}\\
&& -\pa_0\pa_i h^{0 i}-\pa_k\pa_i h^{k j}-\pa_0\pa_j h^{0 
i}-\pa_k\pa_j h^{k 
i}-\pa_i\pa_j h-\pa_0^2 h^{ ij}+\triangle h^{ij}\nn\\
&&+\d^{ij}(\pa_0^2h^{00}+2\pa_0\pa_m h^{0 m}+\pa_m\pa_n 
h^{mn})-\d_{ij}\Box(h^{00}-\psi)=2\tau^{ij}
.\label{ij}
\eea
In order find a solution of these inhomogeneous equations 
we assume the following ansatz for the components of $h_{\mu\nu}$:
\bea
h^{00}&=&d_3\theta^{0\rho}\theta_\rho^{\ 0}r^2+d_4\theta^{0m}\theta^{0n}x^m 
x^n+d_5\theta^{\a\b}\theta_{\a\b}r^2,\nn\\
h^{ij}&=&d_1\theta^{i\rho}\theta_\rho^{\ j} r^2+d_2\theta^{im}\theta^{jn} 
x^mx^n+d_6\d^{ij}\theta^{\a\b}\theta_{\a\b}r^2+d_7\theta^{\a\b}\theta_{\a\b}
x^i 
x^j\nn\\
&&+d_9(\theta^{i\rho}\theta_\rho^{\ 
n}x^nx^j+\theta^{j\rho}\theta_\rho^{\ 
n}x^nx^i)+d_8\theta^{i\rho}\theta_\rho^{\ l}x^nx^l\d_{ij},\nn\\
h^{0i}&=&d_{10}\theta^{0\rho}\theta_\rho^{\ 
i}r^2+d_{11}\theta^{0m}\theta^{in}x^m 
x^n+d_{12}\theta^{0l}\theta_l^{\ n}x^ix^l\ , 
\eea
where $r^2=\sum_{i=1}^3x^ix^i$ and $d_1,\dots,d_{12}$ are arbitrary constants 
to be determined from 
equations (\ref{00}-\ref{ij}).
Inserting this ansatz into equations (\ref{00}-\ref{ij}) leads to the following 
set of
algebraic equations: 
\bea
d_1-d_9+d_8+6d_6-3d_7&=&\frac{33}{8l^6},\nn\\
4(d_1-d_9+d_8)+3d_2+12d_6-6d_7&=&\frac{11}{4l^6},\nn\\
d_1-d_9+d_8+3d_2&=&-\frac{11}{2l^6},\nn\\
d_1-d_9+d_8+d_4&=&-\frac{11}{2l^6},\nn\\
d_1-d_9+d_8+8d_6-4d_7+4d_5-2d_3-d_4&=&\frac{11}{2l^6},\nn\\
d_1-d_9+d_8+4d_6-2d_7+2d_5&=&0,\nn\\
4d_{10}+3d_{11}-2d_{12}&=&-\frac{11}{l^6}
 . \label{JednZaKonstante}
\eea
To start with, let us assume that both $\theta^{0i}$ and $\theta^{ij}$ are
different from zero, $\theta^{0i}\ne0$ and $\theta^{ij}\ne0$. Then
the solution of the previous set of equations is: 
\bea 
&& d_2=-\frac{11}{6l^6}, d_4=-\frac{11}{2l^6}, d_5=-\frac{11}{8l^6}, d_3=0,\nn\\
&& d_1-d_9+d_8=0, \label{konstante}\\
&& d_{10}=-\frac{11}{4l^6}-\frac{3d_{11}}{4}+\frac{d_{12}}{2}, 
d_7=2d_6-\frac{11}{8l^6} .\nn
\eea  
From (\ref{konstante}) it follows that some constants will remain undetermined. 
The presence of undetermined constants suggests the existence of some 
residual symmetry. A detailed analysis of this residual symmetry we postpone 
for future work. In this paper we fix the undetermined constants in the 
following way: $d_1=d_9=d_8=0$, 
$d_{10}=-d_{12}=0$ and
$d_6=-d_7$. Finally, the components of metric tensor follow: 
\bea
g_{00}&=& 1 - \frac{11}{2l^6}\theta^{0m}\theta^{0n}x^m 
x^n-\frac{11}{8l^6}\theta^{\a\b}\theta_{\a\b}r^2,\nn\\
g^{0i} &=& 
-\frac{11}{
3l^6}\theta^{0m}\theta^{in}x^m 
x^n ,\nn\\ 
g_{ij}&=& -\d_{ij} -\frac{11}{6l^6}\theta^{im}\theta^{jn} 
x^mx^n+\frac{11}{24l^6}\d^{ij}\theta^{\a\b}\theta_{\a\b}r^2-\frac{11}{24l^6}
\theta^{\a\b}
\theta_{\a\b}x^i x^j. \label{NCMinkowskiMetric}
\eea  
From the equation (\ref{EoM-metric}) it follows that the scalar curvature of 
the NC Minkowski space-time\footnote{Note that this result is unique and it 
does not depend neither on the way we choose the ansatz for solving the 
equation (\ref{EoM-metric}) in the case of Minkowski space-time not on the 
way we fix the undetermined constants.} is given by
\begin{equation}
R=-\frac{11}{l^6}\theta^{\a\b}\theta^{\g\d}\eta_{\a\g}\eta_{\b\d} =const. 
\label{RNCMikowski}
\end{equation}
This shows that the noncommutativity induces curvature. The 
sign of the scalar curvature will depend on the particular 
values of the
parameter
$\theta^{\a\b}$. For example, if $\theta^{ij}=0$ and $\theta^{0i}\neq 0$ then 
the scalar curvature $R$ is positive. On the other hand, if $\theta^{ij} \neq 
0$ and $\theta^{0i} = 0$ then 
the scalar curvature $R$ is negative. The induced curvature is very small, 
being quadratic in 
$\theta^{\a\b}$ and it
will be difficult to measure it. However, qualitatively we showed that 
noncommutativity is a
source of curvature, just like matter fields or the cosmological constant.

\noindent The Reimann tensor for this solution can be calculated easily. A very 
interesting (and unexpected)
observation follows: knowing the components of the Riemann tensor, the 
components of the metric
tensor can be written as
\bea
g_{00}&=&1-R_{0m0n}x^mx^n,\nn\\ 
g_{0i}&=&-\frac23R_{0min}x^mx^n,\nn\\ 
g_{ij}&=&-\d_{ij}-\frac13R_{imjn}x^mx^n .\label{FermiNCMinkowski}
\eea
This shows that the coordinates $x^\m$ we started with, are Fermi normal 
coordinates. These
coordinates are inertial coordinates of a local observer moving along a
geodesic. The time coordinate $x^0$ is just the proper time of the observer 
moving along the geodesic.
The space coordinates $x^i$ are defined as afine parameters along the geodesics 
in the hypersurface
orthogonal the actual geodesic of the observer. Unlike Riemann normal 
coordinates which can be
constructed in
a small neighborhood of a point, Fermi normal coordinates can
be constructed in a small neighborhood of a geodesic, that is inside a small 
cylinder
surrounding the geodesic \cite{FermiCoordiantes}. Along the geodesic 
these 
coordinates are inertial,
that is 
\begin{equation}
g_{\mu\nu}|_{geod.} = \eta_{\mu\nu}, \quad  \partial_\rho g_{\mu\nu}|_{geod.} = 
0 \, . \nn
\end{equation}
The measurements performed by
the local observer moving along the geodesic are described in the Fermi normal 
coordinates.
Especially, she/he is the one that
measures $\theta^{\alpha\beta}$ to be constant! In any other reference frame 
(any other coordinate
system) observers will
measure $\theta^{\a\b}$ different from constant. 

\initiate
\section{Conclusions}

In this paper we constructed a NC gravity model based on the $SO(2,3)_\star$ 
gauge symmetry. We
used the $\star$-product and the enveloping algebra approach and the SW map. An
effective NC gravity action was constructed using the expansion in the small NC 
parameter $\theta^{\a\b}$. The zeroth order
of the action is the commutative action (\ref{KomDejstvo}). The first order 
correction vanishes. The
second order correction is calculated; the calculation and the result are 
long and cumbersome.
Therefore, we chose to analyze the model sector by sector. In this paper we 
were interested in the low
energy sector, presumable describing physics at low curvatures. In that case 
the action is given by
(\ref{action-linearnoRT}). The equations of motion show that, just like ordinary 
matter,
noncommutativity plays a role of a source for curvature and/or torsion. More 
explicitly, in the
example of NC Minkowski space time, we explicitly calculated the curvature 
induced by
noncommutativity and showed that in the presence of noncommutativity Minkowski 
space-time becomes
curved with a constant scalar curvature. 

In addition we gain a better understanding of the diffeomorphisim symmetry 
breaking problem in the $\theta$-constant NC space-time. Namely, the 
$\theta$-constant deformation is naturally defined for an inertial observer. 
Therefore, it is not possible to apply the $\theta$-constant deformation for GR 
solutions in arbitrary coordinates. With this observation we now understand the 
breaking of diffeomorphism symmetry in
the following way: there is a preferred
reference system defined by the Fermi normal coordinates and the NC parameter 
$\theta^{\a\b}$
is constant in that particular reference system. In an arbitrary reference 
system the NC deformation is obtained by an appropriate coordinate 
transformation. We conclude that the constant NC deformation is consistent only 
with the reference system given by the Fermi normal coordinates. In 
our future work we plan to investigate other solutions of our NC gravity
model, such as the NC Schwartzschild solution and cosmological solutions. 
Especially, 
we are interested
in the role of Fermi normal
coordinates in these solutions and in this way we hope to gain a better 
understanding of NC gravity. Also, using the advantage of the NC gauge theory 
approach we plan to include matter fields in our analysis and see the 
consequences of the noncommutativity on the matter part of the gravity action.

\vskip1cm \noindent 
{\bf Acknowledgement}
\hskip0.3cm
We would like to thank Ilija Simonovi\'c, Milutin Blagojevi\' c,  Maja Buri\' c, 
Dragoljub Go\v canin
and Nikola Konjik for
fruitful discussion and useful comments. The work is
supported by project
ON171031 of the Serbian Ministry of Education and Science and partially
supported  by  the  Action  MP1405  QSPACE from  the  Europe
an  Cooperation  in  Science
and  Technology  (COST).

\appendix

\renewcommand{\theequation}{\Alph{section}.\arabic{equation}}
\initiate
\section{Notation} 

\begin{itemize}

\item {\bf Metric conventions}

\noindent We use the "mostly minus" or West Cost metric convention. The 
Minkowski metric is then
\begin{equation}
{\rm d}s^2 =  {\rm d}t^2 -{\rm d}x^2 -{\rm d}y^2 -{\rm d}z^2 .\label{Mink-}
\end{equation}
The Reimann tensor is defined as:
\be
R_{\m\n\ \l}^{\ \ \rho}=R_{\m\n}^{\ \ ab}e_{a}^\rho e_{\l b}= 
\pa_\m\G^\r_{\n\l} 
-\pa_\n\G^\r_{\m\l}+\G^\r_{\m\kappa}\G^\kappa_{\n\l} 
-\G^\r_{\n\kappa}\G^\kappa_{\m\l} .\label{Riemann}
\ee
Ricci tensor is defined as the following contraction of the Riemann tensor
\be 
R_{\n\l}=R_{\m\n\ \l}^{\ \ \m} .\label{Ricci}
\ee
Then the scalar curvature follows as:
\be 
R=g^{\m\n}R_{\m\n} .\label{R}
\ee
Note that with these conventions the scalar curvature of the vacuum solution 
with the positive cosmological constant (dS space)
is negative, while the scalar curvature of the vacuum solution with the 
negative cosmological solution (AdS space) is positive.

\item {\bf AdS algebra and the $\gamma$-matrices}

\noindent Algebra relations\footnote{$\epsilon^{01235}=+1,\ \epsilon^{0123}=1$}:
\bea 
\{M_{AB},\Gamma_C\}&=&i\epsilon_{ABCDE}M^{DE},\nn\\
\{M_{AB},M_{CD}\}&=&\frac{i}{2}\epsilon_{ABCDE}\Gamma^{E}+\frac12(\eta_{AC}\eta_
{BD}-\eta_{AD}\eta_{BC}),\nn\\
{[}M_{AB},\Gamma_C{]}&=&i(\eta_{BC}\Gamma_A-\eta_{AC}\Gamma_B),\nn\\
\Gamma_A^\dagger&=&-\gamma_0\Gamma_A\gamma_0,\nn\\
M_{AB}^\dagger&=&\gamma_0M_{AB}\gamma_0. \label{AdSrelations}
\eea
Identities with traces:
\bea
&&\tr (\Gamma_A\Gamma_B)=4\eta_{AB},\nn\\
&&\tr (\Gamma_A)=\tr (\Gamma_A\Gamma_B\Gamma_C)=0,\nn\\
&&\tr
(\Gamma_A\Gamma_B\Gamma_C\Gamma_D)=4(\eta_{AB}\eta_{CD}-\eta_{AC}\eta_{BD}+\eta_
{AD}\eta_{CB}),\nn\\
&&\tr (\Gamma_A\Gamma_B\Gamma_C\Gamma_D\Gamma_E)=-4i\epsilon_{ABCDE}\nn\\
&&\tr (M_{AB}M_{CD}\Gamma_E)=i\epsilon_{ABCDE},\nn\\
&&\tr (M_{AB}M_{CD})=-\eta_{AD}\eta_{CB}+\eta_{AC}\eta_{BD} .\label{AdStraces}
\eea

\end{itemize}

\initiate
\section{Expanding actions: useful formulae}

\begin{itemize}

\item {\bf Some results used in calcualtions of $S_{iNC}^{(1)}$ with $i=1,2,3$}:
\begin{eqnarray}
(\hat\phi\star \hat F_{\mu \nu})^{(1)} &=& -\frac{1}{4}\theta^{\alpha 
\beta}\{\o_{\alpha},(\partial_{\beta}+D_{\beta})(\phi F_{\mu 
\nu})\}\nn\\
&& +\frac{i}{2}\theta^{\alpha \beta}D_{\alpha}\phi D_{\beta}F_{\mu 
\nu}+\frac{1}{2}\theta^{\alpha \beta}\phi\{F_{\mu \alpha},F_{\nu \beta}\} 
,\label{B1}
\end{eqnarray}
\begin{eqnarray}
(\hat D_{\rho}\hat\phi\star\hat 
D_{\sigma}\hat\phi)^{(1)}&=&-\frac{1}{4}\theta^{\o 
\beta}\{A_{\alpha},(\partial_{\beta}+D_{\beta})(D_{\rho}\phi 
D_{\sigma}\phi)\}\nn\\
&& +\frac{i}{2}\theta^{\alpha 
\beta}(D_{\alpha}D_{\rho}\phi)(D_{\beta}D{\sigma}\phi)+\frac{1}{2}\theta^{
\alpha \beta}\{F_{\alpha 
\rho},D_{\beta}\phi\}D_{\sigma}\phi\nn\\
&& +\frac{1}{2}\theta^{\alpha 
\beta}D_{\rho}\phi\{F_{\alpha \sigma},D_{\beta}\phi\}. \label{B2}
\end{eqnarray}
Under the intergal, using the partial integration, terms of the form 
$\{A_{\alpha},(\partial_{\beta}+D_{\beta})Y\}$, where the field $Y$ transforms 
in the adjoint representation $\delta_\epsilon Y = i[\epsilon, Y]$ can be 
covariantized:
\be
\int {\rm d}^{4}x\,\theta^{\alpha 
\beta}\tr(\{A_{\alpha},(\partial_{\beta}+D_{\beta})Y\})
=\int {\rm d}^{4}x\, \theta^{\alpha \beta}\tr(\{F_{\alpha \beta},Y\}).\label{B3}
\ee

\item {\bf Some results used in calculations of $S_{iNC}^{(2)}$}:
\begin{eqnarray}
((\hat D_{\alpha}D_{\rho}\hat\phi)\star(\hat D_{\beta}\hat
D_{\sigma}\hat\phi))^{(1)}&=&-\frac{1}{4}\theta^{\gamma
\delta}\{\o_{\gamma},(\partial_{\delta}+D_{\delta})((D_{\alpha}D_{\rho}\phi)(D_{
\beta}D_{\sigma}
\phi))\}\nn\\
&& +\frac{i}{2}\theta^{\gamma
\delta}D_{\gamma}(D_{\alpha}D_{\rho}\phi)D_{\delta}(D_{\beta}
D_{\sigma}\phi)\nn\\
&&+\frac{1}{2}\theta^{\gamma \delta}\{F_{\gamma
\alpha},(D_{\delta}D_{\rho}\phi)\}D_{\beta}D_{\sigma}\phi\nn\\
&&+\frac{1}{2}\theta^{\gamma\delta}D_{\alpha}(\{F_{\gamma
\rho},D_{\delta}\phi\})(D_{\beta}D_{\sigma}\phi)\nn\\
&&+\frac{1}{2}\theta^{\gamma
\delta}(D_{\alpha}D_{\rho}\phi)\{F_{\gamma 
\beta},(D_{\delta}D_{\sigma}\phi)\}\nonumber\\
&& +\frac{1}{2}\theta^{\gamma 
\delta}(D_{\alpha}D_{\rho}\phi)D_{\beta}(\{F_{\gamma
\sigma},D_{\delta}\phi\}), \label{B4}
\end{eqnarray}
\begin{eqnarray}
(\hat{D}_{\rho}\hat{\phi} \star
\hat{D}_{\sigma}\hat{\phi}\star \hat{\phi})^{(1)}&=&
-\frac14\theta^{\a\b}\{\o_\a,D_\b(D_\rho\phi D_\s\phi\phi)\} \label{B7}\\
&&+\frac{i}{2}\theta^{\a\b}D_\a(D_\rho\phi D_\s\phi)D_\b\phi
+\frac12\theta^{\a\b}D_\r\phi 
\{F_{\a\s},D_\b \phi\}\phi \nn\\
&& +\frac12\theta^{\a\b}\{F_{\a\r},D_\b \phi\} +\frac{i}{2}\theta^{\a\b}(D_\a 
D_\r \phi)(D_\b D_\n \phi)\phi. \nn
\end{eqnarray}
The product of (\ref{B1}) and (\ref{B4}) in the first order is given by  
\begin{eqnarray}
&&\Big( \hat\phi\star\hat F_{\mu \nu}\star(\hat D_{\alpha}\hat 
D_{\rho}\hat\phi)\star(\hat D_{\beta}\hat
D_{\sigma}\hat \phi)\Big)^{(1)} = \Big( \hat\phi\star\hat F_{\mu \nu} 
\Big)^{(1)}
(D_{\alpha} D_{\rho}\phi)( D_{\beta} D_{\sigma} \phi)\nn\\
&&+ \,\phi F_{\mu \nu}\phi\Big( (\hat D_{\alpha}D_{\rho}\hat\phi)\star(\hat 
D_{\beta}\hat
D_{\sigma}\hat\phi)\Big)^{(1)}+\frac{i}{2}\theta^{\g\d}\pa_\g(\phi 
F_{\m\n})\pa_\d(D_\a D_\r\phi D_\b
D_\s\phi)\nn\\
&& = -\frac{1}{4}\theta^{\gamma 
\delta}\{\omega_{\gamma},(\partial_{\delta}+D_{\delta})(\phi F_{\mu
\nu} D_{\alpha}D_{\rho}\phi D_{\b}D_{\s}\phi)\}\nn\\
&&\hspace*{0.4cm}+\frac{i}{2}\theta^{\g\d}D_\g(\phi F_{\m\n})D_\d( 
D_{\alpha}D_{\rho}\phi
D_{\b}D_{\s}\phi)\nn\\
&&\hspace*{0.4cm}+(\frac{i}{2}D_{\gamma}\phi D_{\delta}F_{\mu 
\nu}+\frac{1}{2}\phi\{F_{\mu \gamma},F_{\nu
\delta}\})(D_{\alpha}D_{\rho}\phi)(D_{\beta}D_{\sigma}\phi)\nonumber\\
&&\hspace*{0.4cm}+\phi F_{\mu
\nu}(\frac{i}{2}D_{\gamma}(D_{\alpha}D_{\rho}\phi)D_{\delta}(D_{\beta}D_{\sigma}
\phi)+\frac{1}{2}\{
F_{\gamma \alpha},(D_{\delta}D_{\rho}\phi)\}(D_{\beta}D_{\sigma}\phi)\nonumber\\
&&\hspace*{0.4cm}+\frac{1}{2}D_{\alpha}(\{F_{\gamma
\rho},D_{\delta}\phi\})(D_{\beta}D_{\sigma}\phi)+\frac{1}{2}(D_{\alpha}D_{\rho}
\phi)\{F_{\gamma
\beta},(D_{\delta}D_{\sigma}\phi)\}\nn\\
&&\hspace*{0.4cm}+\frac{1}{2}(D_{\alpha}D_{\rho}\phi)D_{\beta}(\{F_{\gamma 
\sigma},D_{\delta}\phi\}))).
\label{B5}
\end{eqnarray}
The noncovariant term in (\ref{B5}) under the integral is rewritten using 
(\ref{B3}). Then the
fully covariant form of the expanded term (\ref{B5}) is obtained:
\begin{eqnarray}
&&\Big( \hat\phi\star\hat F_{\mu \nu}\star(\hat D_{\alpha}\hat 
D_{\rho}\hat\phi)\star(\hat D_{\beta}\hat
D_{\sigma}\hat \phi)\Big)^{(1)}=
\frac{i}{4}\theta^{\alpha \beta}\theta^{\gamma \delta}\int {\rm d}^{4}x\, 
\epsilon^{\mu \nu \rho
\sigma}\tr \Big\{\nn\\
&&-\frac{1}{4}\{F_{\gamma \delta},\phi F_{\mu
\nu}(D_{\alpha}D_{\rho}\phi)(D_{\beta}D_{\sigma}\phi)\}
+\frac{i}{2}D_{\gamma}(\phi F_{\mu
\nu})D_{\delta}((D_{\alpha}D_{\rho}\phi)(D_{\beta}D_{\sigma}\phi))\nn\\
&&+(\frac{i}{2}D_{\gamma}\phi
D_{\delta}F_{\mu \nu}+\frac{1}{2}\phi\{F_{\mu \gamma},F_{\nu
\delta}\})(D_{\alpha}D_{\rho}\phi)(D_{\beta}D_{\sigma}\phi)\nonumber\\
&&+\phi F_{\mu
\nu}(\frac{i}{2}D_{\gamma}(D_{\alpha}D_{\rho}\phi)D_{\delta}(D_{\beta}D_{\sigma}
\phi)+\frac{1}{2}\{
F_{\gamma 
\alpha},(D_{\delta}D_{\rho}\phi)\}(D_{\beta}D_{\sigma}\phi)+\nonumber\\
&&+\frac{1}{2}D_{\alpha}(\{F_{\gamma
\rho},D_{\delta}\phi\})(D_{\beta}D_{\sigma}\phi)+\frac{1}{2}(D_{\alpha}D_{\rho}
\phi)\{F_{\gamma
\beta},(D_{\delta}D_{\sigma}\phi)\}\nn\\&&+\frac{1}{2}(D_{\alpha}D_{\rho}\phi)D_
{\beta}(\{F_{\gamma
\sigma},D_{\delta}\phi\}))\Big\}. \label{B6}
\end{eqnarray}
The remaining terms in (\ref{S_2-drugi-red-1}) are calculated following the same 
steps.

\end{itemize}

\initiate
\section{Variation of the action $S_{NC}$}

Here we write some useful formulas for calculating equarions of motion:
\begin{eqnarray}
\delta_\omega R_{\a\b}^{ab} &=& \nabla_\a\delta\omega_\b^{ab} - 
\nabla_\b\delta\omega_\a^{ab} ,\label{C1}\\ 
\delta_e R_{\a\b}^{ab} &=& 0,\label{C2}\\
\delta_\omega T_{\a\b}^{a} &=& -e_{\a b}\delta\omega_\b^{ab} + 
e_{\b b}\delta\omega_\a^{ab} ,\label{C3}\\ 
\delta_e T_{\a\b}^{a} &=& \nabla_\a\delta e_\b^a - \nabla_\b\delta e_\a^a 
,\label{C4}\\
\delta_e e &=& ee_a^\m\delta e_\m^a ,\quad \delta_e g_{\rho\sigma} = 
(\delta_\rho^\m e_{\sigma a} + \delta_\sigma^\m e_{\rho a})\delta e_\m^a 
.\label{C5}
\end{eqnarray}


\begin{thebibliography}{99}

\bibitem{GRWaves}
B. P. Abbott et al. (LIGO Scientific Collaboration and Virgo Collaboration), {\it Observation of
Gravitational Waves from a Binary Black Hole Merger}, Phys.\ Rev.\ Lett.\ {\bf 116}, 061102
(2016), [arXiv:1602.03837].

B. P. Abbott et al. (LIGO Scientific Collaboration and Virgo Collaboration), {\it GW151226:
Observation of Gravitational Waves from a 22-Solar-Mass Binary Black Hole Coalescence}
Phys.\ Rev.\ Lett. {\bf 116}, 241103 (2016).

\bibitem{CERN}
ATLAS Collaboration, {\it Search for resonances in diphoton events at
$\sqrt{s} =13 TeV$ with the ATLAS detector}, arXiv:1606.03833.

CMS Collaboration, {\it Search for resonant production of high-mass photon 
pairs in proton-proton collisions at $\sqrt{s} =8$ and $13 TeV$}, 
arXiv:1606.04093.

ATLAS and CMS Collaborations, {\it Combined Measurement of the Higgs Boson 
Mass in pp Collisions at $\sqrt{s}=7$ and $8 TeV$ with the ATLAS and CMS 
Experiments}, Phys.\ Rev.\ Lett.\ {\bf 114}, 33 (2015), [arXiv:1503.07589]. 


\bibitem{TwistApproach}
P.Aschieri, C. Blohmann, M. Dimitrijevi\' c, F. Meyer, P. Schupp
and J. Wess, {\it A Gravity Theory on Noncommutative Spaces},
Class.\ Quant.\ Grav. {\bf 22}, 3511 (2005), [hep-th/0504183 ].

P. Aschieri, M. Dimitrijevi\' c, F. Meyer and  J. Wess,
{\it Noncommutative Geometry and Gravity}, Class.\ Quant.\ Grav. {\bf
23}, 1883 (2006), [hep-th/0510059].

\bibitem{TwistSolutions}
T. Ohl and A. Schenckel, {\it  	
Cosmological and Black Hole Spacetimes in Twisted Noncommutative Gravity}, JHEP {\bf 0910} (2009)
052, [arXiv: 0906.2730].

P. Aschieri and L. Castellani, {\it Noncommutative Gravity Solutions}, J.\ 
Geom.\ Phys.\ {\bf 60}, 375-393 (2010), [arXiv:0906.2774].

\bibitem{Chaichian}
M. Chaichian, P.P. Kulish, K. Nishijima, A. Tureanu, {\it On a 
Lorentz-invariant interpretation of noncommutative space-time and its 
implications on noncommutative QFT}, Physics Letters B 604 1-2 (2004), 
[hep-th/0408069].

M. Chaichian, P. Pres̃najder, A. Tureanu, {\it New Concept of Relativistic 
Invariance in Noncommutative Space-
Time: Twisted Poincaré Symmetry and Its Implications}, Phys. Rev. Lett. 94, 
151602 (2005) [hep-th/0409096].


\bibitem{EmGravityApproach} 
H. S. Yang, {\it Emergent gravity from noncommutative spacetime}, Int.\ J.\
Mod.\ Phys. {\bf A24}, 4473 (2009), [hep-th/0611174].

H. Steinacker, {\it Emergent Geometry and Gravity from Matrix Models: an
Introduction}, Class.\ Quant.\ Grav.\ {\bf 27}, 133001
(2010), [arXiv:1003.4134].

\bibitem{OtherApproaches}
M. Buri\' c and J. Madore, {\it Spherically Symmetric Noncommutative Space: $d =
4$}, Eur.\ Phys.\ J. {\bf C58},  347 (2008), [arXiv: 0807.0960].

M. Buri\' c and J. Madore, {\it On noncommutative spherically symmetric spaces},
arXiv:1401.3652.

L. Tomassini, S. Viaggiu, {\it Building non-commutative spacetimes at the Planck length for
Friedmann flat cosmologies}, Class.\ Quant.\ Grav. {\bf 31} 185001
(2014), [arXiv:1308.2767].

\bibitem{SWmapApproach} A. H. Chamseeddine, {\it Deforming Einstein's gravity},
Phys.\ Lett.\ B {\bf 504} 33 (2001), [hep-th/0009153].

M. A. Cardella and D. Zanon,
{\it Noncommutative deformation of four-dimensional gravity}, Class.\ Quant.\
Grav. {\bf 20}, L95 (2003), [hep-th/0212071].

\bibitem{PL09}
P. Aschieri and L. Castellani, {\it Noncommutative $D=4$ gravity
coupled to fermions} JHEP, {\bf 0906}, 086 (2009), [arXiv:0902.3823].

\bibitem{PLSUGRA}
P. Aschieri and L. Castellani, {\it Noncommutative supergravity in $D=3$ and 
$D=4$}, JHEP {\bf 0906}, 087 (2009), [arXiv:0902.3823].

L. Castellani, {\it Chern-Simons supergravities, with a twist}, JHEP {\bf 
1307}, 133 (2013), [arXiv:1305.1566].

\bibitem{Dobrski}
M. Dobrski, {\it Background independent noncommutative gravity from Fedosov 
quantization of endomorphism bundle}, arXiv:1512.04504.

M. Dobrski, {\it On some models of geometric noncommutative general 
relativity}, Phys.\ Rev.\ D {\bf 84}, 065005 (2011), [arXiv:1011.0165].

\bibitem{Ostali}
A. Kobakhidze, C. Lagger and A. Manning, {\it Constraining noncommutative 
spacetime from GW150914}, Phys.\ Rev.\ D {\bf 94} 064033 (2016), 
[arXiv:1607.03776],

D. Klammer and H. Steinacker, {\it Cosmological solutions of emergent 
noncommutative gravity}, Phys.\ Rev.\ Lett.\ {\bf 102} (2009) 221301, 
[arXiv:0903.0986]. 

E. Harikumar and V. O. Rivelles, {\it Noncommutative Gravity}, Class.\ Quant.\ 
Grav.\ {\bf 23}, 7551-7560 (2006), [hep-th/0607115].

\bibitem{MajaJohn}
M. Buri\'c, T. Grammatikopoulos, J. Madore, G. Zoupanos, {\it Gravity and the 
Structure of
Noncommutative Algebras}, JHEP {\bf 0604} 054, 2006, [hep-th/0603044].

M. Buri\'c, J. Madore, G. Zoupanos, {\it The Energy-momentum of a Poisson 
structure},
Eur.\ Phys.\ J.\ {\bf C 55} 489-498, 2008, [arXiv:0709.3159].

\bibitem{UsLetter}
M. Dimitrijevi\' c \'Ciri\'c, B. Nikoli\'c and V. Radovanovi\' c, {\it 
Noncommutative gravity and the relevance of the $\theta$-constant deformation}, 
 arXiv:1609.06469.

\bibitem{MiAdSGrav}
M. Dimitrijevi\' c, V. Radovanovi\' c and H. \v Stefan\v ci\' c,
{\it AdS-inspired noncommutative gravity on the Moyal plane},
Phys. Rev. D {\bf 86}, 105041 (2012), [arXiv:1207.4675].

\bibitem{MDVR-14}
M. Dimitrijevi\' c and  V. Radovanovi\' c,
{\it Noncommutative $SO(2,3) $ gauge theory and noncommutative gravity},
Phys. Rev. D {\bf 89}, 125021 (2014), [arXiv:1404.4213].

\bibitem{Wilczek}
F. Wilczek, {\it  	
Riemann-Einstein structure from volume and gauge symmetry}, Phys.\ Rev.\ Lett.\ 
{\bf 80} (1998)
4851-4854, [hep-th/9801184].


\bibitem{stelle-west} K. S. Stelle and P. C. West, {\it Spontaneously
broken de Sitter symmetry and the gravitational holonomy group}, Phys.\ Rev D
{\bf 21}, 1466 (1980).

S. W. MacDowell and F. Mansouri,
{\it Unified geometrical theory of gravity and supergravity}, Phys.\ Rev.\ Lett.
{\bf 38}, 739 (1977).

P. K. Towsend, {\it Small-scale structure of spacetime as
the origin of the gravitation constant}, Phys.\ Rev.\ D {\bf 15},  2795 (1977).

\bibitem{SWMapEnvAlgebra}
B. Jur\v{c}o, L. M\"oller, S.~Schraml, P.~Schupp and J.~Wess,
{\it Construction of non-Abelian gauge theories on noncommutative spaces},
Eur.\ Phys.\ J.\ C{\bf 21}, 383 (2001), [hep-th/0104153].

N.~Seiberg and E.~Witten,
{\it String theory and noncommutative geometry},
JHEP {\bf 09}, 032 (1999), [hep-th/9908142].




\bibitem{FermiCoordiantes}
F.K. Manasse and C.W. Misner, {\it Fermi Normal Coordinates and Some Basic Concepts in
Differential Geometry}, J.\ Math.\ Phys.\ {\bf 4} (1963) 735-745.

C. Chicone and B. Mashoon, {\it Explicit Fermi coordinates and tidal dynamics in de Sitter and Godel
spacetimes}, Phys.\ Rev.\ {\bf D 74} (2006) 064019, [gr-qc/0511129]. 

D. Klein and E. Randles, {\it  	
Fermi coordinates, simultaneity, and expanding space in Robertson-Walker cosmologies}, Annales Henri
Poincare 12 (2011) 303-328, [arXiv:1010.0588].







\end{thebibliography}
\end{document}